\documentclass[conference]{IEEEtran}
\IEEEoverridecommandlockouts
\usepackage{cite}
\usepackage{amsmath,amssymb,amsfonts}
\usepackage{algorithmic}
\usepackage{graphicx}
\usepackage{textcomp}
\usepackage{xcolor}
\usepackage{comment}
\usepackage{multirow}
\usepackage{subcaption}
\usepackage{booktabs}
\usepackage{pdfpages}
\def\BibTeX{{\rm B\kern-.05em{\sc i\kern-.025em b}\kern-.08em
    T\kern-.1667em\lower.7ex\hbox{E}\kern-.125emX}}
\begin{document}

\onecolumn
\begin{center}\Huge {Preface to the arXiv Edition}\\\vspace{1cm}\end{center}
\textit{ This document is an extended version of our paper accepted at ISIT 2026. It is organized into two parts: 
\begin{itemize}
        \item \textbf{Part I (Conference Version)} contains the accepted conference version, focusing on the implementation and provides a comprehensive experimental evaluation, including coding efficiency comparisons against RaptorQ (the state-of-the-art rateless code) and LT codes and decoding speed metrics compared to RaptorQ. We explicitly note three specific updates and clarifications in Part I:
        \begin{enumerate}
        \item Regarding the comparison with LT codes, we update the block size threshold: its coding efficiency does not match that of METTLE until the block size grows to approximately $k = 500,000$.
        \item The overhead ratios mentioned in the evaluation all explicitly account for \textit{tail loss}.
        \item We did not explicitly mention the "left-to-right peeling" effect in the text, although this property is implicitly demonstrated by the low latency results shown.
        \end{enumerate}
        \item \textbf{Part II (Supplementary Material)} provides the foundational description of the METTLE coding scheme, elaborating on its design choices, favorable properties, and the tail compression technique. It also includes a comparison to packet-level LDPC and an analysis of coding efficiency under time-varying Gilbert-Elliott (GE) channels versus BEC channels to demonstrate the scheme's resilience. Note while discussion in Part II remain factually valid, the evaluation was preliminary, and the implementation details were presented at a high level.
        \item \textbf{Reading Guide:} We recommend readers start with \textbf{Part I} for the most up-to-date performance benchmarks. Readers interested in the evolution of the scheme design and the detailed discussion should refer to \textbf{Part II}.
\end{itemize}
}
\begin{center}
    \vspace*{1in}
    \Huge \textbf{Part I} \\
    \vspace{1cm}
\end{center}

%%%%%%%%%% Part I %%%%%%%%%%
\title{METTLE: Efficient Streaming Erasure Code with Peeling Decodability}
\thanks{A provisional patent application has been filed covering the METTLE scheme and implementation described in this work~\cite{mettle_patent_2026}.}
\author{\IEEEauthorblockN{Qianru Yu}
\IEEEauthorblockA{
%\textit{College of Computing} \\
\textit{Georgia Tech}\\
%Atlanta, USA \\
qyu87@gatech.edu}
\and
\IEEEauthorblockN{Tianji Yang}
\IEEEauthorblockA{
%\textit{College of Computing} \\
\textit{Georgia Tech}\\
%Atlanta, USA \\
tyang425@gatech.edu}
\and
\IEEEauthorblockN{Jingfan Meng}
\IEEEauthorblockA{
%\textit{College of Computing} \\
\textit{Georgia Tech}\\
%Atlanta, USA \\
jeffmeng@live.com}
\and
\IEEEauthorblockN{Jun (Jim) Xu}
\IEEEauthorblockA{
%\textit{College of Computing} \\
\textit{Georgia Tech}\\
%Atlanta, USA \\
jx@cc.gatech.edu}
}
\twocolumn
\maketitle
\begin{abstract}
In this work, we solve a long-standing open problem in coding theory with broad applications in networking and systems: designing an erasure code that simultaneously satisfies three requirements -- (1) high coding efficiency, (2) low coding complexity, and (3) being a streaming code (defined as one with low decoding latency).  
We propose METTLE (Multi-Edge Type with Touch-less Leading Edge), the first erasure code to meet all three requirements. Compared to ``streaming RaptorQ" (RaptorQ configured with a small source block size to ensure a low decoding latency), METTLE is only slightly worse in coding efficiency, but 47.7 to 84.6 times faster to decode.
\end{abstract}

\section{Introduction}\label{sec:intro}
In this paper, we solve a long-standing open problem in coding theory that has many applications in networking and systems.
The open problem is to design a streaming erasure code (defined as one that has a low decoding latency) that has (1) high coding efficiency and (2) low coding complexity.  As we will elaborate next, no existing streaming (erasure) code possesses both properties.
For example, all existing streaming codes~\cite{martinian2004burst,martinian2007delay,badr2013streaming,badr2016layered,Domanovitz2022streaming,rudow2022streaming} are based on Maximum distance separable (MDS) codes~\cite{Singleton1964maximum} such as Reed-Solomon (RS)~\cite{Reed_Solomon_codes}, with a small block size (e.g., $n=96$ symbols in~\cite{rudow2023tambur}).  
Although RS is optimal in coding efficiency, it has a high decoding complexity:  even when using such a small block size, its decoding speed can support an (real-time interactive video) application throughput of at most 15 Mbps~\cite{rudow2023tambur}.   

Fountain codes, including Tornado~\cite{luby1997practical} and its successor LT (Luby Transform)~\cite{luby_lt_2002}, can arguably serve as streaming codes when a small source block size $k$ is used (to ensure low decoding latency). 
Their decoding complexities are also quite low: to decode each codeword symbol (packet), Tornado (resp. LT) only needs to perform $O(1)$ (resp. $O(\log k)$) peeling operations,
each equivalent to a belief propagation (BP) message update under BEC.
Hence, such codes are at least two orders of magnitude faster than RS-based streaming codes (e.g., convolutional RS~\cite{martinian2004burst, martinian2007delay}) to decode, even when $k$ is small. 
However, when $k$ is small, their coding efficiencies are quite low.   
For example, when $k = 400$ (packets), the coding overhead of LT needs to be around 75\% even when there is zero erasure.

Raptor codes~\cite{shokrollahi_raptor_2006} were proposed to improve this coding efficiency.  To this end, Raptor codes augment LT with an LDPC+HDPC precoding.  
RaptorQ~\cite{rfc6330}, the ``best of the Raptors,'' can achieve near-optimal coding efficiency:  the decoder needs to receive only $k+1$ codeword symbols to decode for the $k$ source symbols (or one symbol away from being optimal).  
However, to achieve such high coding efficiency, RaptorQ must resort to Gaussian elimination for the LDPC+HDPC decoding, and as a result pays a high decoding complexity cost: RaptorQ
is only a few times faster than RS to decode~\cite{libraptorq}.

In this work, we propose METTLE (Multi-Edge Type with Touch-less Leading Edge), a novel erasure code that solves this open problem.  
METTLE is a streaming code that supports on-the-fly decoding.  
Other merits of METTLE are best illustrated through a comparison with RaptorQ, when both are viewed as solutions to the low coding-efficiency problem of LT code under small $k$.
RaptorQ trades high decoding complexity for near-optimal coding efficiency, as explained above.
This has been widely accepted as the best attainable compromise for nearly two decades~\cite{shokrollahi_raptor_2006, shokrollahi2011raptor}.
In contrast, METTLE achieves a markedly better tradeoff:  while METTLE's overall coding efficiency remains only slightly worse than that of RaptorQ, METTLE decodes between 47.7$\times$ and 84.6$\times$ faster than RaptorQ due to its purely peeling-based decoding.
This translates into around 5 Gbps for METTLE, and no more than 100 Mbps for RaptorQ, in decoding throughput on a single CPU core.

Furthermore, METTLE exhibits an even more favorable tradeoff than RaptorQ in the large-$k$ regime. 
As $k$ increases beyond a few hundreds (of symbols), RaptorQ's coding efficiency improves only marginally, whereas its decoding complexity (and time) per symbol scales superlinearly. 
For example, as reported in~\cite{libraptorq}, when $k=27,000$, RaptorQ's decoding time per symbol (1500-byte packet) exceeds 130ms, whereas METTLE requires only 2.6~$\mu$s! 

Beyond solving this open problem in coding theory, METTLE offers three other attractive properties.  First, METTLE (which is non-systematic in this paper) has a systematic variant with the same coding efficiency~\cite{archive}. 
Second, METTLE has impressive mettle:  it is highly resilient to erasures with time-varying rates and burst erasures (Section~\ref{subsec:coding_efficiency}).  

Third, METTLE is a hashing-based code, in the sense its Tanner graph~\cite{richardson2008modern} is defined implicitly via hashing.
METTLE derives the following three important (computer) ``systems benefits'' from that. 
First, METTLE's coding operation does not require storage of its Tanner graph in memory, making it nearly stateless and lightweight to implement.   
Second, METTLE is a \textit{continuous rate} code in the sense that it can be configured to use any desired coding overhead ratio, unlike $(n, k)$ block codes whose overhead ratio is fixed to $(n-k)/k$ by design.  
Third, METTLE can vary this overhead ratio on-the-fly in response to changing channel conditions, without incurring much (network) protocol processing overheads. In a sense, METTLE enables seamless, continuous rate adaptation.
\section{METTLE Coding Scheme}\label{sec:mettle_scheme}
In coding theory terms, METTLE is an SC-MET-LDGM (Spatially-Coupled Multi-Edge-Type Low-Density Generator Matrix) code. METTLE is also deeply rooted in theoretical computer science (TCS):
METTLE is a novel, nontrivial (explained in~\cite{archive}) adaptation of a hashing-based data structure (HBDS), which we call Walzer's HBDS~\cite{walzer_code}, into a streaming erasure code (so its Tanner graph is hash-generated).  
Walzer's HBDS was originally designed to improve the compactness (space efficiency) of an HBDS called invertible Bloom lookup table (IBLT)~\cite{IBLT}, using {\it spatial coupling} (SC)~\cite{felstrom1999time,Urbanke_spatial_coupling,kudekar2011threshold,kudekar2013capacity}.   

Our presentation follows a natural progression from IBLT (Section~\ref{subsec:iblt}) to Walzer's HBDS (Section~\ref{subsec:walzer_scheme}), and finally to METTLE (Section~\ref{subsec:mettle}).
In the rest of the paper, for notational cleanliness, we omit floor and ceiling operators even when they would otherwise be required.

\subsection{IBLT (Invertible Bloom Lookup Table)}\label{subsec:iblt}

An IBLT data structure~\cite{Eppstein_WhatsDifference_2011}, typically used for encoding a set $\mathcal{S}$, can be viewed as an LDGM codeword containing $(1+c)k$ symbols, where $c>0$ corresponds roughly to its coding overhead ratio and $k\approx |\mathcal{S}|$. 
Each codeword symbol is considered a {\it bin} whose value is the bitwise-XOR of the {\it balls} (source symbols) that are hash-thrown into the bin.  

IBLT's encoding works as follows.  Each element (source symbol) in $\mathcal{S}$ is considered a ball.  
The encoding process throws all these $|\mathcal{S}|$ balls, each ball being thrown a constant number (say $l$) of times, into these bins.  
Each ball is thrown into $l$ bins whose indices are i.i.d. random variables (RVs) uniformly distributed in the IBLT's (array) index range $[0, (1+c)k)$, 
determined by $l$ different hash functions $h_i$, $i=1, \cdots, l$.  

IBLT decodes via an iterative peeling process.  In each iteration, it locates all bins containing exactly one ball.  
Each such ball is thus ``exposed" (with its value say $x$ revealed) and removed (peeled) from all $l-1$ other bins that the ball $x$ is thrown into.  
The peeling operations in the current iteration may cause more balls to be exposed and peeled in the next iteration, which propels the peeling process. 

Neither IBLT, nor its original target application---set reconciliation~\cite{Dodis_PinSketch_2008, Eppstein_WhatsDifference_2011, yang-rateless-iblt}---is typically regarded as part of the error correction coding (ECC) literature by the HBDS research community.
Nevertheless, BIFF code~\cite{Mitzenmacher2012Biff,Mitzenmacher2012objectreconciliation} demonstrates that IBLT can in fact be used for error correction.  In BIFF, the discrepancy between a transmitted file $S$ (viewed as a set) and the received file $S'$, after passing through a lossy and noisy channel, is corrected using an IBLT constructed over $S$ that is transmitted alongside the data.
There are two key differences between BIFF and METTLE. First, BIFF is a block code, whereas METTLE is a streaming (blockless) code. Second, BIFF builds directly on IBLT, whereas METTLE introduces significant and consequential modifications to Walzer's IBLT.

%%%%%%%%%%%%%%%%%%%%%%%%%%%%% WALZER %%%%%%%%%%%%%%%%%%%%%%%%%%%%%%%%
%\vspace{-2pt}
\subsection{Walzer's HBDS}\label{subsec:walzer_scheme}
Like in IBLT, Walzer's HBDS encodes approximately $k$ balls, but now consists of~$(1+c)(k+w)$ bins, where~$w>0$ denotes the (spatial-) {\it coupling window} size.  
A key difference between IBLT and Walzer's HBDS lies in the range of the $l$ (independent) uniform hash functions~$h_i(\cdot)$, $1\le i\le l$.
In the IBLT, each~$h_i$ maps uniformly over the entire index range~$[0, (1+c)k)$, meaning any ball~$x$ can land anywhere in the data structure.
In contrast, Walzer's HBDS localizes each~$h_i(x)$ to a small neighborhood of width~$(1+c)w$ in its index range $[0, (1+c)k)$, as follows.
This ball $x$ is first mapped to a uniform random {\it position}~$\xi$ in the ``ball range"~$[0, k)$ using a different uniform hash function than $h_j(\cdot)$, $1\le j\le l$. 
Then, for each $i$, the hash value $h_i(x)$ is uniformly distributed in the range $[(1+c)\xi, (1+c)(\xi+w))$, or in other words {\it coupled} to the ball position $\xi$. 

\begin{figure}[htbp]
  \centering
    \includegraphics[width=0.95\linewidth]{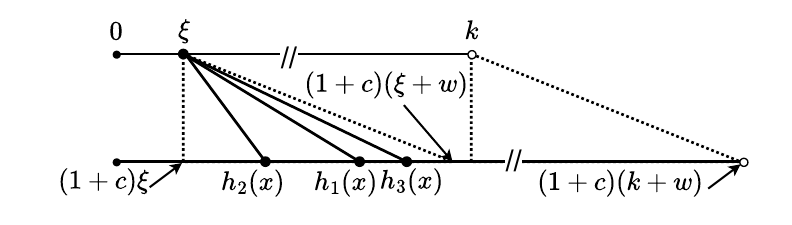}
    \caption{Spatial coupling in Walzer's HBDS.}
    \label{fig:scheme1}
\end{figure}

This {\it spatially coupled} hashing in Walzer's HBDS is illustrated in Fig.~\ref{fig:scheme1}. The top line represents the ``ball range" $[0, k)$, and the bottom line represents the ``bin range" $[0,(1+c)(k+w))$.  To visualize the concept of spatial coupling, the bottom line is scaled by a factor of~$1/(1+c)$, so that a ball at position~$\xi$ on the top line aligns vertically with the bin at position~$(1+c)\xi$ on the bottom line.  
We use $l = 3$ hash function in all figures (to avoid ``overcrowding" them).  
In practice, $l$ is typically between 3 and 5 in both Walzer's HBDS and METTLE.

%%%%%%%%%%%%%%%%%%%%%%%%%%%%% METTLE %%%%%%%%%%%%%%%%%%%%%%%%%%%%%%%%

\subsection{METTLE}\label{subsec:mettle}
METTLE adapts Walzer's HBDS into a streaming code, through three major modifications:  (1) from spatial coupling to ``time coupling," (2) TLE (touch-less leading edge), and (3) MET (multi-edge type).  

\subsubsection{Baseline: from spatial coupling to ``time coupling"} \label{subsubsec:time_coupling}

METTLE inherits Walzer's spatially coupled IBLT data structure, but converts its space coupling to “time coupling.” 
Consider a block of $k$ time-ordered packets where $k$ can be arbitrarily large and may not be known in advance. In METTLE, the encoding of these packets works as follows.  
Upon the arrival of the first packet (ball), the encoder assigns to it the (spatial) ball position $0$ (on the top line in Fig.~\ref{fig:scheme1}), and accordingly throws it into $l$ bins independently and uniformly distributed in the bin range $[0, (1+c)w)$ (that is spatially coupled to its ball position $0$).  
Similarly, the second packet is placed at ball position $1$ and thrown into $l$ uniform random bins in the bin range $[1+c, (1+c)(1+w))$ (that is spatially coupled to its ball position $1$), and so on. 
In other words, the position of a ball is no longer determined by hashing as in Walzer's HBDS but by the time order (sequence number) of the ball.
Hence, we refer to $w$ as the \textit{time-coupling window} instead.  We call this version (without MET and TLE) the baseline,
whose coding efficiency is not great (so MET and TLE are indispensable), as shown in~\cite{archive}.

\subsubsection{MET (Multi-Edge Type) modification}\label{subsubsec:MET}

Recall that in the baseline, a ball at position $x$ (corresponding to the $(x+1)^{\text{th}}$ uncoded packet) is thrown into
$l$ bins whose indices $h_i(x)$, $i = 1, 2, \cdots, l$, are RVs that are i.i.d. uniform over the window $[(1+c)x, (1+c)(x+w))$ time-coupled with (the ball position) $x$.
The MET (Multi-Edge Type) modification is to make these $l$ RVs independent but not identically distributed.  

\begin{figure}[htbp]
  \centering
    \includegraphics[width=0.9\linewidth]{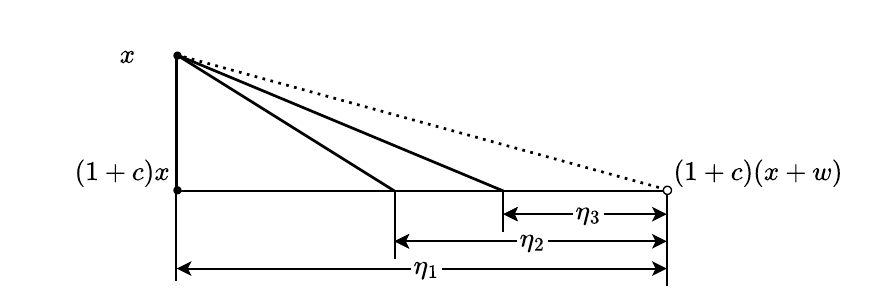}
    \caption{Illustration of MET coupling.}
    \label{fig:scheme2}
\end{figure}

To simplify the notations going forward, we replace each hash value (notation) $h_i(x)$, 
with an RV $\eta_i \triangleq (1+c)(x+w) - h_i(x)$, that captures the same information, for $i = 1, 2, \cdots, l$.
Fig.~\ref{fig:scheme2} zooms into the ``MET-modified'' time-coupling window $[(1+c)x, (1+c)(x+w))$, using this new $\eta$ notation (with $l = 3$).
As shown in Fig.~\ref{fig:scheme2}, each $\eta_i$, $i = 1, 2, \cdots, l$, is the distance, from the window's right boundary, to the ``landing position" of the $i^{\text{th}}$ edge fanning out from the ``ball'' at position $x$.
In the baseline, these $l$ RVs are i.i.d. uniform in this range. 

The name MET comes from the following fact:  since these $l$ edges now have distinct (landing position) distributions, they have different types in the density evolution (DE) analysis~\cite{richardson2008modern, richardson2002multi}.
The final landing-position distributions, optimized (for coding efficiency) via numerical DE computations, verified by simulations, and adopted in METTLE are as follows:  $\eta_i$ follows $\mathrm{Binomial}((1+c)w, 1/2^{i-1})$, for $i = 2, 3, \cdots, l$.  
As such, the landing positions corresponding to $\eta_2$, $\eta_3$, $\cdots$, $\eta_l$ respectively become roughly exponentially closer to $(1+c)(x+w)$, the right boundary of the time-coupling window.

Spatial coupling (SC) takes a very different form in SC-LDPC than in METTLE.
In SC-LDPC, coupling occurs between nearby variable and check vertices, where each variable vertex (resp. check vertex) represents a large sub-block of variable bits (resp. check bits). The coupling width---defined as the maximum number of positions (sub-blocks) separating a coupled variable-check vertex pair---is typically no more than 5.
In contrast, in METTLE, each corresponding vertex represents a single source or codeword symbol (packet), and the coupling width used in this work ranges from 600 to 1000.
In addition, multi-edge-type (MET) structure is rarely used in spatially coupled LDPC or LDGM codes, as it offers limited benefit beyond that already provided by spatial coupling. Even when MET is incorporated---such as in the spatially-coupled MacKay-Neal code and Hsu-Anastasopoulos code ensembles (which are hybrid LDPC-LDGM constructions)~\cite{obata2013scmet}---the MET connectivity is confined within each variable-check sub-block pair. Moreover, for the multiple trunk edges emanating from a check sub-block (to variable sub-blocks within the coupling window), the associated MET structures are i.i.d. and uniformly distributed.
In contrast, in METTLE, all $l$ edges emanating from each ball are coupling (degenerate) trunk edges, and these edges are of distinct types.
These differences are fundamental: unlike SC-LDPC and its MET variants, which rely on block-level graph constructions with local and largely i.i.d. edge-type assignments on trunk edges, METTLE operates via symbol-level coupling with globally structured, heterogeneous edge types. Consequently, METTLE lies outside the standard SC-MET-LDPC framework.

\subsubsection{TLE (Touch-less Leading Edge) modification}

The TLE modification concerns $\eta_1$, the landing position of the first edge.  
The TLE technique is to simply let $\eta_1 = (1+c)w$; as such, the RV $\eta_1$ becomes a constant. 
In this case, the ``first edge'' becomes deterministic and is vertically incident on the bin located at the left boundary of its time-coupling window, 
as shown in Fig.~\ref{fig:scheme2}.
In other words, for the ball at each position $y$ ($y = 0, 1, 2, \cdots, n-1$), we define $h_1(y) = (1+c)y$ (now we ``need the first hash function back''), meaning that the ``first edge'' fanning out from the ball at position $y$ is always incident on the bin $(1+c)y$.  
Note that $h_1(\cdot)$ is injective, or the ``first edges'' fanning out from any two distinct balls will be incident on different bins, because two neighboring incident positions (bins) have distance $1+c>1$ between them. We say these ``first edges'' are touch-less because no two of them will touch each other (or fall into the same bin).  

\subsubsection{Overhead ratio, window size, and tail loss}

METTLE is not a block code and can use an arbitrarily large source block size $k$ without sacrificing decoding latency.  
This latency, however, grows with the window size $w$, so we keep $w$ below 1000. 
When $k$ is large,  
the overhead ratio is close to~$c$.  
The tail (rate) loss $(1+c)w$ due to SC termination~\cite{mitchell2015spatially} can be cut by half to $(1+c)w/2$ 
using a tail compression technique~\cite{archive}, which we use and account for in METTLE evaluation.

\subsubsection{Capacity-approaching in the large-$w$ regime}

We proved in~\cite{archive} that the baseline METTLE is capacity-approaching over the BEC, provided that the window size $w$ grows sufficiently large (along with $k$) 
so that the density evolution (DE) analysis within a window remains accurate. In this large-$w$ regime, however, the decoding latency may become too large for the baseline METTLE to qualify as a streaming code. This result extends Walzer's capacity-approaching proof, which does not account for bin erasures (as such erasures do not arise in the intended applications of Walzer's HBDS).

Empirically, the full METTLE approaches capacity faster than the baseline METTLE in this large-$w$ regime; that is, it exhibits better finite-length asymptotics. Somewhat ironically, however, we currently lack a rigorous proof that METTLE itself is capacity-approaching. The difficulty stems from the fact that the mathematical techniques used in Walzer's proof do not extend to METTLE. Specifically, due to the use of MET and TLE, the DE equations of METTLE take the form of a vector (functional) recursion, rather than the scalar recursion that arises in Walzer's HBDS.
Existing theoretical frameworks for coupled vector recursions~\cite{Yelda2012vector, obata2013scmet} are limited, and in particular require symmetry conditions that METTLE's DE equations provably do not satisfy~\cite{Yelda2012vector}.

\section{METTLE Performance Evaluation}\label{sec:eval}

Our evaluation results are summarized as follows.  First, Section~\ref{subsec:decoding_latency} shows that METTLE's average decoding latency ranges from 37 to 199 symbols (1500-byte packets), roughly 18--95~ms at the 4K streaming bitrate of 25~Mbps, under a wide range of (memoryless) BEC and Gilbert-Elliott (GE) channels.  
Second, Section~\ref{subsec:coding_efficiency} shows that METTLE's coding efficiency is slightly worse than RaptorQ's, but much better than LT's. Third, Section~\ref{subsec:decoding_speed} shows that METTLE's decoding time per symbol is 2.6~$\mu$s, which is 47.7 to 84.6 times faster than RaptorQ's, even when RaptorQ uses a small source block size.

\subsection{BEC and Gilbert-Elliott (GE) configurations used}

A memoryless BEC (Binary Erasure Channel) is parameterized by the parameter $\varepsilon$, the probability each symbol (packet) is erased. We denote it as BEC($\varepsilon$) when convenient. 
Five $\varepsilon$ values are used in our evaluation:  1\%, 2\%, 3\%, 8\%, and 10\%. 

A Gilbert-Elliott (GE) channel is a two-state (``good" and ``bad") Markov-modulated (hence not memoryless) BEC~\cite{yang2014markov-modulated-bec}.   
that is described by four parameters:  $\epsilon_{g}, \epsilon_{b}, P_{g2 b}, P_{b2 g}$.
For each packet entering the channel, the channel erases it with probability $\epsilon_{g}$ (resp. $\epsilon_{b}$) and transitions to the ``bad" state (resp. the ``good'' state) with probability $P_{g2b}$ (resp. $P_{b2g}$), 
if the channel is currently in the ``good" state (resp. the ``bad'' state).
The average erasure probability of a GE channel is $P_{avg} = (P_{g2 b}+P_{b2 g})^{-1}(\epsilon_{b}P_{g2 b}+\epsilon_{g}P_{b2 g})$. 
When $\epsilon_{b}=1$, the channel experiences bursts of consecutive (packet) erasures. 

\begin{table}[htbp]
  \caption{GE (Gilbert-Elliott) channels.}
  \label{tab:ge_params}
  \begin{tabular}{cccccc}
    \toprule
    Channel&$P_{g2 b}$&$P_{b2 g}$&$\epsilon_{g}$&$\epsilon_{b}$&$\epsilon_{\text{avg}}$\\
    \midrule
    VoIP (GE1)& $5\times 10^{-4}$ & 0.2 
    &0.01&1&1.25\%\\
    WiMAX (GE2)& 0.04 & 0.05 &0.01&0.02&1.44\%\\
    Video-conf-light (GE3)& 0.05 & 0.75 &0.01&0.1&1.56\%\\
    Video-conf-heavy (GE4)& 0.05 & 0.75 &0.05&0.5&7.81\%\\
    Long-fade (GE5)& 0.001 & 0.01 &0.01&0.1&1.82\%\\
  \bottomrule
\end{tabular}
\end{table}

For our evaluation, we use five GE channels summarized in Table~\ref{tab:ge_params}.
The first four were previously used in~\cite{badr2017fec,al2013intra,rudow2023tambur} to model a variety of network application dynamics.
Their parameters, as reported in~\cite{badr2017fec,al2013intra,rudow2023tambur}, were obtained by fitting the corresponding application traces to the two-state GE model.
Specifically, VoIP~\cite{badr2017fec}, aliased as GE1 for spacing considerations, models bursty erasures observed in a VoIP trace, with a mean burst length of 5 in the ``bad" state.
WiMAX~\cite{al2013intra} (GE2) models erasures encountered by a video streaming application over a WiMAX (wireless) link.
Video-conf-light (GE3) and Video-conf-heavy (GE4), both from~\cite{rudow2023tambur}, model erasures observed in real-world videoconferencing traces.

Long-fade (GE5), the last channel in Table~\ref{tab:ge_params}, is created by us and models a slow-fading~\cite{Sklar1997rayleigh} wireless channel in which the ``bad" state persists for an average of 100 packets, with
$\epsilon_{b} = 0.1$.  Despite its moderate average erasure rate of 1.82\%, this channel constitutes a particularly stringent stress test for erasure codes due to its extremely long bad-state duration.

\subsection{Decoding latency of METTLE}\label{subsec:decoding_latency}

Although both LT and RaptorQ are rateless codes, they operate over a fixed source block (of size $k$) and therefore incur block-level decoding latency. Consequently, their average decoding latency is approximately $k/2$ symbols, with a maximum latency of $k$ symbols.  
In contrast, METTLE is fully blockless and does not incur such block-level latency.  Instead, its decoding latency---both average and 95th percentile---is measured empirically at the symbol level across individual experiments.

The same parameter settings, described next, are used across all three evaluations of METTLE---decoding latency, coding efficiency, and decoding speed---to ensure consistency and fairness.
We use $l = 4$ edges per source symbol $x$, with exponentially decaying landing positions, following the construction illustrated in Fig.~\ref{fig:scheme1} (which depicts the case $l = 3$).
The coupling window size $w$ is set to $600$, as our experiments over $w\in [400, 1000]$ show that $w=600$ strikes a good trade-off between coding efficiency, which decreases with increasing $w$, and decoding latency, which increases with increasing $w$.

Table~\ref{tab:latency1} reports the average decoding latencies of METTLE under five BECs, abbreviated by their erasure probabilities (e.g., BEC(0.01) abbreviated as 0.01), and five GE configurations, abbreviated by their aliases listed in Table~\ref{tab:ge_params}.
We make two observations from these results.
First, most decoding latencies are small in absolute terms, with the largest being 199 packets (each 1500 bytes), corresponding to approximately 95.5~ms even at a 4K streaming bitrate of 25 Mbps.  By comparison, end-to-end latencies of up to 150~ms are generally acceptable for real-time applications~\cite{ITU-TG1010}.
Second, decoding latency generally increases with the erasure rate, characterized by $\epsilon$ for BECs and $\epsilon_{avg}$ for GE channels.
For each BEC or GE configuration, METTLE uses a fixed coding overhead that is not inflated to reduce decoding latency; this overhead is instead chosen in the next 
subsection solely to meet a target decoding success probability.

\begin{table}[htbp]
  \centering
  \caption{METTLE's average decoding latency.}
  \label{tab:latency1}
  \setlength{\tabcolsep}{4pt}
  \begin{tabular}{cccccccccc}
  \toprule
  0.01 & 0.02 & 0.03 & 0.08 & 0.1 & GE1 & GE2 & GE3 & GE4 & GE5\\
  \midrule
  61 & 84 & 117 & 133 & 199 & 37 & 72 & 53 & 127 & 50\\
  \bottomrule
\end{tabular}
\end{table}
\vspace{-1em}
\begin{table}[htbp]
  \centering
  \caption{METTLE's 95th percentile decoding latency.}
  \label{tab:latency2}
  \setlength{\tabcolsep}{4pt}
  \begin{tabular}{cccccccccc}
  \toprule
  0.01 & 0.02 & 0.03 & 0.08 & 0.1 & GE1 & GE2 & GE3 & GE4 & GE5\\
  \midrule
  459 & 582 & 663 & 728 & 815 & 291 & 522 & 425 & 713 & 420 \\
  \bottomrule
\end{tabular}
\end{table}

Arguably, a streaming code should provide stricter, rather than merely average, decoding-latency guarantees~\cite{rudow2022streaming}. 
Accordingly, we also measure METTLE's 95th-percentile decoding latency under the same ten channel conditions.  As shown in Table~\ref{tab:latency2}, the resulting latencies 
range from 291 to 815 packets, corresponding to roughly 150--400~ms at a 4K video bitrate; these values are admittedly above the conventional 150~ms real-time threshold for today's 4K streaming. 
However, the same packet-level latencies translate to one to two orders of magnitude smaller wall-clock delays for emerging applications---such as immersive XR (AR/VR/MR), 8K--12K video, and telesurgery---that routinely operate at data rates of hundreds of Mbps to multiple Gbps~\cite{apple_vision_pro_2023,super_hivision_2012,telesurgery}.

\subsection{Coding efficiency comparison}\label{subsec:coding_efficiency}

In this section, we compare METTLE with RaptorQ and LT in coding efficiency.
In this evaluation, this efficiency is measured by the overhead ratio required to achieve a target decoding failure probability of $10^{-3}$.

For RaptorQ and LT, each decoding failure means that the codeword encoding a source block (of size $k$, which is tens to hundreds as we will elaborate shortly) fails to decode.   
For METTLE, we use a stricter decoding success definition, at our disadvantage.
A METTLE decoding failure refers to the situation that the number of erasures is large enough to cause the leftward peeling process to stall somewhere along a ``mega-codeword'' of size $N=10^5(1+c)$ packets, so that the source symbols beyond the stall position to be mostly unrecoverable.  
The corresponding decoding success definition is stricter, because absent such as a stall, all source symbols in the mega-codeword (rather than a small codeword of size $k(1+c)$ for RaptorQ and LT), except for an isolated few that may ``hit the error floor'' (elaborated in~\cite{archive}) under heavy erasure, can be decoded. 
Note that a METTLE decoding failure is inexpensive to cure: the encoder, upon the request of the decoder, only needs to retransmit several erased symbols around the stall to get the peeling process going again.

\begin{table}[htbp]
\centering
\caption{Coding efficiency comparison.}
\label{tab:coding_eff}
\begin{tabular}{c|cccc}
\toprule
\textbf{Channel}&\textbf{METTLE}&\textbf{RaptorQ} &\textbf{LT} \\
\midrule
BEC(0.01)&5.5\%& 6.14\% ($k$=114)&76\% ($k$=400) \\
BEC(0.02)&8\%& 7.14\% ($k$=168)&78\% ($k$=400) \\
BEC(0.03)&9\%& 7.63\% ($k$=236)&81\% ($k$=400) \\
BEC(0.08)&20\%& 15.6\% ($k$=269)&92\% ($k$=400) \\
BEC(0.1)&25\%&15\% ($k$=405)&96\% ($k$=400) \\
VoIP&9\%&23.8\% ($k$=84)&76\% ($k$=400) \\
WiMAX&6\%&6.04\% ($k$=149)&78\% ($k$=400) \\
Video-conf-light&8\%&7.02\% ($k$=114)&78\% ($k$=400)\\
Video-conf-heavy&20\%&15.56\% ($k$=257)&93\% ($k$=400)\\
Long-fade&12\%&15.84\% ($k$=101)&78\% ($k$=400) \\
\bottomrule
\end{tabular}
\end{table}
\vspace{-1em}

Table~\ref{tab:coding_eff} reports the coding overhead ratio $c$ required, for METTLE, RaptorQ, and LT, to achieve $10^{-3}$ decoding failure probability, under the ten BEC and GE channels.
For each channel, we configure the source block size $k$ of RaptorQ to be approximately $2\tau$, where $\tau$ denotes METTLE's decoding latency under the same channel.  (The approximation arises because RaptorQ supports only a discrete set of $(n,k)$ block designs due to its precoding structure.) This choice ensures that METTLE's decoding latency is comparable to that of RaptorQ, whose expected decoding latency is $k/2$, as explained earlier. For LT, we fix $k = 400$ across all ten channels, which is roughly twice METTLE's maximum observed decoding latency (199 packets) among these channels, thereby biasing the comparison against METTLE.

We first focus on the comparison between METTLE and RaptorQ.
Under the five BECs, METTLE has either better or slightly worse coding efficiency than RaptorQ when the erasure rate is small ($\varepsilon = 0.01, 0.02, 0.03$), and has 
worse coding efficiency when the erasure rate is large ($\varepsilon= 0.08, 0.1$).  
Under the five GE channels, METTLE has better coding efficiency under VoIP and Long-fade, worse efficiency under Video-conf-heavy, and similar efficiencies under WiMAX and Video-conf-light.
As mentioned earlier, the erasures in VoIP and Long-fade are the most bursty among the five GE channels.  METTLE's outperformance under them confirms our earlier claim
that METTLE is particularly resilient under bursty erasures.
METTLE is also far more efficient than LT, whose $c$ ranges from 76\% to 96\% (despite that $k = 400$).  

We also compare METTLE with RaptorQ and LT when keeping their 95th-percentile decoding latencies similar. 
Our experiments show that this alternative criterion does not help RaptorQ and LT much.
For example, when $k=1002$, Raptor's coding efficiency under BEC(0.1) improves only slightly to $c=14.87\%$ (from $c=15\%$), but its decoding speed becomes $237$ times slower than METTLE's.

\subsection{Decoding speed}\label{subsec:decoding_speed}

In this section, we compare the decoding speeds of METTLE and RaptorQ. For fairness, both decoders are implemented in C++ and ran on the same workstation.
For RaptorQ, we use a widely adopted C++ implementation from~\cite{libraptorq}.
The metric for this comparison is decoding time per packet.

METTLE's decoding time is only 2.6~$\mu$s per packet.  
Even when $k$ is between $100$ and $400$ (for the latency-matched coding efficiency comparison), RaptorQ's decoding time per packet is between 124 and 220~$\mu$s,
which is between 47.7$\times$ and 84.6$\times$ longer than METTLE's.  
When $k$ grows larger than $511$, the decoding time per packet increases superlinearly, as shown in Table~\ref{tab:decoding_speed}.
For example, when $k = 1002$ in the example above, RaptorQ's decoding time per packet is 616~$\mu$s, which is 237 times longer than METTLE's.

\begin{table}[htbp]
\centering
\caption{Decoding time \emph{per source symbol} for RaptorQ.}
\label{tab:decoding_speed}
\begin{tabular}{c|ccccccc}
\toprule
$k$&127&257&511&1002&2040&4069&8194\\
\midrule
Time ($\mu$s) &122&150&265&616&3545&5824&21451\\
\bottomrule
\end{tabular}
\end{table}

\section{Conclusion}
In this work, we present METTLE, a novel streaming erasure code with both high coding efficiency and low decoding complexity.
Compared to RaptorQ (with a small $k$), METTLE is 47.7$\times$ to 84.6$\times$ faster to decode, yet achieves only slightly worse coding efficiency. 

\section*{Acknowledgment}
This work was supported in part by the National Science Foundation under Grant No. CNS-2007006 and by a seed gift from Dolby Laboratories.

\bibliographystyle{IEEEtran}
\bibliography{reference-arXiv}

@String{Computing = "Computing" }

@String{Computer = "{IEEE} Computer" }

@String{Academic = "Academic Press" }

@String{Springer = "Springer-Verlag" }

@inproceedings{luby_lt_2002,
	address = {Vancouver, BC, Canada},
	title = {{LT} codes},
	isbn = {978-0-7695-1822-0},
	url = {http://ieeexplore.ieee.org/document/1181950/},
	doi = {10.1109/SFCS.2002.1181950},
	urldate = {2024-09-16},
	booktitle = {The 43rd {Annual} {IEEE} {Symposium} on {Foundations} of {Computer} {Science}, 2002. {Proceedings}.},
	publisher = {IEEE Comput. Soc},
	author = {Luby, M.},
	year = {2002},
	pages = {271--280}
}

@inproceedings{luby1997practical,
author = {Luby, Michael G. and Mitzenmacher, Michael and Shokrollahi, M. Amin and Spielman, Daniel A. and Stemann, Volker},
title = {Practical loss-resilient codes},
year = {1997},
isbn = {0897918886},
publisher = {Association for Computing Machinery},
address = {New York, NY, USA},
url = {https://doi.org/10.1145/258533.258573},
doi = {10.1145/258533.258573},
booktitle = {Proceedings of the Twenty-Ninth Annual ACM Symposium on Theory of Computing},
pages = {150-159},
numpages = {10},
location = {El Paso, Texas, USA},
series = {STOC '97}
}

@article{walzer_code,
author = {Walzer, Stefan},
title = {Peeling Close to the Orientability Threshold Spatial Coupling in Hashing-Based Data Structures},
year = {2025},
issue_date = {July 2025},
publisher = {Association for Computing Machinery},
address = {New York, NY, USA},
volume = {21},
number = {3},
issn = {1549-6325},
url = {https://doi.org/10.1145/3711822},
doi = {10.1145/3711822},
journal = {ACM Trans. Algorithms},
month = jul,
articleno = {33},
numpages = {23}
}

@ARTICLE{Urbanke_spatial_coupling,
  author={Kudekar, Shrinivas and Richardson, Thomas J. and Urbanke, Rüdiger L.},
  journal={IEEE Transactions on Information Theory}, 
  title={Wave-Like Solutions of General 1-D Spatially Coupled Systems}, 
  year={2015},
  volume={61},
  number={8},
  pages={4117-4157},
  doi={10.1109/TIT.2015.2438870}
}

@article{Reed_Solomon_codes,
author = {Reed, I. S. and Solomon, G.},
title = {Polynomial Codes Over Certain Finite Fields},
journal = {Journal of the Society for Industrial and Applied Mathematics},
volume = {8},
number = {2},
pages = {300-304},
year = {1960},
doi = {10.1137/0108018},
URL = {https://doi.org/10.1137/0108018}
}

@INPROCEEDINGS{IBLT,
  author={Goodrich, Michael T. and Mitzenmacher, Michael},
  booktitle={2011 49th Annual Allerton Conference on Communication, Control, and Computing (Allerton)}, 
  title={Invertible bloom lookup tables}, 
  year={2011},
  publisher = {IEEE},
  address= {Monticello, IL, USA},
  volume={},
  number={},
  pages={792-799},
  doi={10.1109/Allerton.2011.6120248}}

@ARTICLE{shokrollahi_raptor_2006,
  author={Shokrollahi, Amin},
  journal={IEEE Transactions on Information Theory}, 
  title={Raptor codes}, 
  year={2006},
  volume={52},
  number={6},
  pages={2551-2567},
  doi={10.1109/TIT.2006.874390}}

@article{rudow2022streaming,
  title={Streaming codes for variable-size messages},
  author={Rudow, Michael and Rashmi, KV},
  journal={IEEE Transactions on Information Theory},
  volume={68},
  number={9},
  pages={5823--5849},
  year={2022},
  publisher={IEEE}
}

@article{martinian2004burst,
  title={Burst erasure correction codes with low decoding delay},
  author={Martinian, Emin and Sundberg, C-EW},
  journal={IEEE Transactions on Information theory},
  volume={50},
  number={10},
  pages={2494--2502},
  year={2004},
  publisher={IEEE}
}

@INPROCEEDINGS{martinian2007delay,
  author={Martinian, Emin and Trott, Mitchell},
  booktitle={2007 IEEE International Symposium on Information Theory}, 
  title={Delay-Optimal Burst Erasure Code Construction}, 
  year={2007},
  publisher = {IEEE},
  address={Nice, France},
  pages={1006-1010},
  doi={10.1109/ISIT.2007.4557355}}

@article{badr2016layered,
  title={Layered constructions for low-delay streaming codes},
  author={Badr, Ahmed and Patil, Pratik and Khisti, Ashish and Tan, Wai-Tian and Apostolopoulos, John},
  journal={IEEE Transactions on Information Theory},
  volume={63},
  number={1},
  pages={111--141},
  year={2016},
  publisher={IEEE}
}

@article{mitchell2015spatially,
  title={Spatially coupled LDPC codes constructed from protographs},
  author={Mitchell, David GM and Lentmaier, Michael and Costello, Daniel J},
  journal={IEEE Transactions on Information Theory},
  volume={61},
  number={9},
  pages={4866--4889},
  year={2015},
  publisher={IEEE}
}

@BOOK{shokrollahi2011raptor,
  title={Raptor codes},
  author={Shokrollahi, Amin and Luby, Michael},
  journal={Foundations and trends{\textregistered} in communications and information theory},
  volume={6},
  number={3--4},
  pages={213--322},
  year={2011},
  publisher={Now Publishers, Inc.},
  doi={10.1561/0100000060}
}

@INPROCEEDINGS{kudekar2011threshold,
  author={Kudekar, Shrinivas and Richardson, Thomas J. and Urbanke, Rüdiger L.},
  booktitle={2010 IEEE International Symposium on Information Theory}, 
  title={{Threshold saturation via spatial coupling: Why convolutional LDPC ensembles perform so well over the BEC}}, 
  year={2010},
  volume={},
  number={},
  pages={684-688},
  doi={10.1109/ISIT.2010.5513587}}

@ARTICLE{Domanovitz2022streaming,
  author={Domanovitz, Elad and Fong, Silas L. and Khisti, Ashish},
  journal={IEEE Transactions on Information Theory}, 
  title={An Explicit Rate-Optimal Streaming Code for Channels With Burst and Arbitrary Erasures}, 
  year={2022},
  volume={68},
  number={1},
  pages={47-65},
  doi={10.1109/TIT.2021.3121101}}

@inproceedings {rudow2023tambur,
author = {Michael Rudow and Francis Y. Yan and Abhishek Kumar and Ganesh Ananthanarayanan and Martin Ellis and K.V. Rashmi},
title = {Tambur: Efficient loss recovery for videoconferencing via streaming codes},
booktitle = {20th USENIX Symposium on Networked Systems Design and Implementation (NSDI 23)},
year = {2023},
isbn = {978-1-939133-33-5},
address = {Boston, MA},
pages = {953--971},
publisher = {USENIX Association},
month = apr
}

@INPROCEEDINGS{Yelda2012vector,
  author={Yedla, Arvind and Jian, Yung-Yih and Nguyen, Phong S. and Pfister, Henry D.},
  booktitle={2012 IEEE Information Theory Workshop}, 
  title={A simple proof of threshold saturation for coupled vector recursions}, 
  year={2012},
  publisher = {IEEE},
  address = {Lausanne, Switzerland},
  volume={},
  number={},
  pages={25-29},
  doi={10.1109/ITW.2012.6404671}}

@INPROCEEDINGS{obata2013scmet,
  author={Obata, Naruomi and Jian, Yung-Yih and Kasai, Kenta and Pfister, Henry D.},
  booktitle={2013 IEEE International Symposium on Information Theory}, 
  title={Spatially-coupled multi-edge type LDPC codes with bounded degrees that achieve capacity on the BEC under BP decoding}, 
  year={2013},
  publisher = {IEEE},
  address = {Istanbul, Turkey},
  pages={2433-2437},
  doi={10.1109/ISIT.2013.6620663}}

@book{richardson2008modern,
  author    = {Richardson, Tom and Urbanke, Ruediger},
  title     = {Modern Coding Theory},
  year      = {2008},
  publisher = {Cambridge University Press},
  address   = {Cambridge},
  isbn      = {9780521852296},
  doi       = {10.1017/CBO9780511791338}
}

@techreport{ITU-TG1010,
  author       = {International Telecommunication Union},
  institution  = {ITU-T},
  title        = {{ITU-T} Recommendation {G}.1010: End-user multimedia {QoS} categories},
  number       = {G.1010 (11/2001)},
  month        = nov,
  year         = {2001}
}

@INPROCEEDINGS{badr2017fec,
  author={Badr, Ahmed and Khisti, Ashish and Tan, Wai-tian and Zhu, Xiaoqing and Apostolopoulos, John},
  booktitle={IEEE INFOCOM 2017 - IEEE Conference on Computer Communications}, 
  title={{FEC} for {VoIP} using dual-delay streaming codes}, 
  year={2017},
  publisher = {IEEE},
  address = {Atlanta, GA, USA},
  pages={1-9},
  doi={10.1109/INFOCOM.2017.8057027}}

@article{al2013intra,
  title={Intra-Refresh Provision for {WiMAX} Data-Partitioned Video Streaming},
  author={Al-Jobouri, Laith and Fleury, Martin and Ghanbari, Mohammed},
  journal={Consumer Electronics Times},
  volume={2},
  number={3},
  pages={137--145},
  year={2013},
  publisher={World Academic Publishing}
}

@article{felstrom1999time,
  title={Time-varying periodic convolutional codes with low-density parity-check matrix},
  author={Felstrom, A Jimenez and Zigangirov, Kamil Sh},
  journal={IEEE Transactions on Information Theory},
  volume={45},
  number={6},
  pages={2181--2191},
  year={1999},
  publisher={IEEE}
}

@incollection{telesurgery,
title = {Chapter 15 - Telesurgery applications, current status, and future perspectives in technologies and ethics},
editor = {Stênio de Cássio Zequi and Hongliang Ren},
booktitle = {Handbook of Robotic Surgery},
publisher = {Academic Press},
pages = {161-168},
year = {2025},
isbn = {978-0-443-13271-1},
doi = {https://doi.org/10.1016/B978-0-443-13271-1.00027-3},
author = {Thiago Camelo Mourão and Shady Saikali and Evan Patel and Mischa Dohler and Vipul Patel and Márcio Covas Moschovas}
}

@ARTICLE{Sklar1997rayleigh,
  author={Sklar, B.},
  journal={IEEE Communications Magazine}, 
  title={Rayleigh fading channels in mobile digital communication systems. I. Characterization}, 
  year={1997},
  volume={35},
  number={9},
  pages={136-146},
  doi={10.1109/35.620535}}

@ARTICLE{kudekar2013capacity,
  author={Kudekar, Shrinivas and Richardson, Thomas J. and Urbanke, Rüdiger L.},
  journal={IEEE Transactions on Information Theory}, 
  title={Spatially Coupled Ensembles Universally Achieve Capacity Under Belief Propagation}, 
  year={2013},
  volume={59},
  number={12},
  pages={7761-7813},
  doi={10.1109/TIT.2013.2280915}}

@inproceedings{richardson2002multi,
  title={Multi-edge type {LDPC} codes},
  author={Richardson, Tom and Urbanke, R{\"u}diger},
  year={2002},
  journal={Workshop honoring Prof. Bob McEliece on his 60th birthday},
  pages={24-25},
  address={California Institute of Technology, Pasadena, California}
}

@ARTICLE{Singleton1964maximum,
  author={Singleton, R.},
  journal={IEEE Transactions on Information Theory}, 
  title={{Maximum Distance Q-Nary Codes}}, 
  year={1964},
  volume={10},
  number={2},
  pages={116-118},
  doi={10.1109/TIT.1964.1053661}}

@article{Eppstein_WhatsDifference_2011,
author = {Eppstein, David and Goodrich, Michael T. and Uyeda, Frank and Varghese, George},
title = {What's the difference? efficient set reconciliation without prior context},
year = {2011},
issue_date = {August 2011},
publisher = {Association for Computing Machinery},
address = {New York, NY, USA},
volume = {41},
number = {4},
issn = {0146-4833},
url = {https://doi.org/10.1145/2043164.2018462},
doi = {10.1145/2043164.2018462},
journal = {SIGCOMM Comput. Commun. Rev.},
month = aug,
pages = {218-229},
numpages = {12}
}

@InProceedings{Dodis_PinSketch_2008,
author="Dodis, Yevgeniy
and Reyzin, Leonid
and Smith, Adam",
editor="Cachin, Christian
and Camenisch, Jan L.",
title="Fuzzy Extractors: How to Generate Strong Keys from Biometrics and Other Noisy Data",
booktitle="Advances in Cryptology - EUROCRYPT 2004",
year="2004",
publisher="Springer Berlin Heidelberg",
address="Berlin, Heidelberg",
pages="523--540",
isbn="978-3-540-24676-3"
}

@inproceedings{yang-rateless-iblt,
  author    = {Yang, Lei and Gilad, Yossi and Alizadeh, Mohammad},
  title     = {Practical Rateless Set Reconciliation},
  year      = {2024},
  isbn      = {9798400706141},
  publisher = {Association for Computing Machinery},
  address   = {New York, NY, USA},
  url       = {https://doi.org/10.1145/3651890.3672219},
  doi       = {10.1145/3651890.3672219},
  booktitle = {Proceedings of the ACM SIGCOMM 2024 Conference},
  pages     = {595–612},
  numpages  = {18},
  location  = {Sydney, NSW, Australia},
  series    = {ACM SIGCOMM '24}
}

@ARTICLE{yang2014markov-modulated-bec,
  author={Yang, Yang and Tan, Jian and Shroff, Ness B. and Gamal, Hesham El},
  journal={IEEE Transactions on Information Theory}, 
  title={Delay Asymptotics With Retransmissions and Incremental Redundancy Codes Over Erasure Channels}, 
  year={2014},
  volume={60},
  number={3},
  pages={1932-1944},
  doi={10.1109/TIT.2014.2300485}}

@misc{rfc6330,
  author = {A. Shokrollahi and M. Luby and M. Watson and T. Stockhammer and L. Minder},
  title = {{RaptorQ} Forward Error Correction Scheme for Object Delivery},
  howpublished = {RFC 6330},
  year = {2011},
  institution = {IETF},
  note = {Available at \url{https://datatracker.ietf.org/doc/html/rfc6330}}
}

@INPROCEEDINGS{badr2013streaming,
  author={Badr, Ahmed and Khisti, Ashish and Tan, Wai-Tian and Apostolopoulos, John},
  booktitle={2013 Proceedings IEEE INFOCOM}, 
  title={Streaming codes for channels with burst and isolated erasures}, 
  year={2013},
  publisher = {IEEE},
  address = {Turin, Italy},
  volume={},
  number={},
  pages={2850-2858},
  doi={10.1109/INFCOM.2013.6567095}}

@INPROCEEDINGS{Mitzenmacher2012Biff,
  author={Mitzenmacher, Michael and Varghese, George},
  booktitle={2012 IEEE International Symposium on Information Theory Proceedings}, 
  title={Biff (Bloom filter) codes: Fast error correction for large data sets}, 
  year={2012},
  volume={},
  number={},
  pages={483-487},
  doi={10.1109/ISIT.2012.6284236}}

@INPROCEEDINGS{Mitzenmacher2012objectreconciliation,
  author={Mitzenmacher, Michael and Varghese, George},
  booktitle={2012 50th Annual Allerton Conference on Communication, Control, and Computing (Allerton)}, 
  title={The complexity of object reconciliation, and open problems related to set difference and coding}, 
  year={2012},
  volume={},
  number={},
  pages={1126-1132},
  doi={10.1109/Allerton.2012.6483345}}

@misc{libraptorq,
  title        = {{libRaptorQ: A C++} Implementation of {RaptorQ} Forward Error Correction Codes},
  author       = {Fulchir, Luca},
  howpublished = {\url{https://github.com/LucaFulchir/libRaptorQ}},
  year         = {2013}
}

@article{archive,
  title={{METTLE: Efficient Streaming Erasure Code with Peeling Decodability} ({arXiv} version)},
  author={Yu, Qianru and Yang, Tianji and Meng, Jingfan and Xu, Jun},
  journal={arXiv preprint arXiv:2602.10020},
  year={2026},
  month={Feb},
  archivePrefix={arXiv},
  primaryClass={cs.IT},
}

@misc{apple_vision_pro_2023,
  author       = {{Apple Inc.}},
  title        = {Apple Vision Pro: Spatial Computing},
  howpublished = {\url{https://www.apple.com/apple-vision-pro/}},
  year         = {2023}
}

@article{super_hivision_2012,
  title={{`Super Hi-Vision'} as next-generation television and its video parameters},
  author={Yamashita, Takayuki and Masuda, Hiroyasu and Masaoka, Kenichiro and Ohmura, Kohei and Emoto, Masaki and Nishida, Yukihiro and Sugawara, Masayuki},
  journal={Information Display},
  volume={28},
  number={11-12},
  pages={12--17},
  year={2012},
  publisher={Wiley Online Library}
}

@misc{mettle_patent_2026,
  title        = {{METTLE}: Native Streaming Code with Peeling Decodability},
  author       = {Xu, Jun and Yu, Qianru and Yang, Tianji and Meng, Jingfan}, 
  year         = {2026},
  month        = jan,
  howpublished = {U.S. Provisional Patent Application No. 63/968,434},
  note         = {(Patent Pending)}
}

%%%%%%%%%% Part II %%%%%%%%%%
\clearpage
\onecolumn 
\begin{center}
    \vspace*{3in}
    \Huge \textbf{Part II} \\
    \vspace{1cm}
    \Large \textbf{Supplementary Material} \\
    \vspace{0.5cm}
\end{center}
\newpage



\section*{Supplementary material}

# Part II

Supplementary Material

# METTLE: Native Streaming Code with Peeling Decodability

QIANRU YU, Georgia Tech, USA
TIANJI YANG, Georgia Tech, USA
JINGFAN MENG, Georgia Tech, USA
JUN (JIM) XU, Georgia Tech, USA

In this work, we address a long-standing open problem in coding theory with broad applications in networking and systems: designing an error correction code (ECC) that simultaneously satisfies three requirements—(1) high coding efficiency, (2) low coding complexity, and (3) being a streaming code, meaning each packet can be decoded with low average latency. We propose METTLE (Multi-Edge Type code with Touch-less Leading Edge), the first ECC scheme to meet all three requirements. Like Tornado and Luby Transform (LT) codes, METTLE belongs to the "peeling" family, where encoding and decoding of each packet requires only a small constant number of XOR operations, ensuring inherently low complexity. Also similar to LT, METTLE achieves very high coding efficiency. Unlike LT, however, METTLE does so without sacrificing stream decodability. For example, with the same coding overhead, METTLE provides the erasure correction capability of an LT code with a large block size of 500,000 packets, while achieving an average decoding latency of only 57 packets under 1% packet loss. METTLE lives up to its name: it displays impressive resilience (mettle) against challenging erasure patterns such as burst erasures. Finally, we say METTLE is a "native" streaming code because, unlike other streaming codes, it is not a block code such as Reed-Solomon with an artificially small block size.

## 1 Introduction

In this paper, we solve a long-standing open problem in coding theory that has many applications in networking and systems. In this problem, a large source message (e.g., a streaming video file) is divided into many packets that need to be coded up to protect against packet losses (called *erasures* in the coding literature) that can occur during their transmissions over a wired or wireless network. In coding theory terms, such a lossy channel is called *binary erasure channel* (BEC), each uncoded packet is called a *source symbol*, and each coded packet is called a *codeword symbol*. In most error-correction coding (ECC) schemes, each symbol (whether source or codeword) corresponds to a bit, a byte, or a short word (e.g., 16 bits long). Hence, when each symbol instead corresponds to an entire packet – potentially thousands of bytes long – in an ECC scheme, as we assume throughout this work, the scheme is said to *operate at the packet level*. Throughout this paper, we will use the terms *symbols* and *packets* interchangeably, and strategically (usually to match the context).

This open problem is to design an ECC scheme that simultaneously satisfies the following three requirements. The first requirement is high coding efficiency – that is, using as little overhead as possible for protection. The second requirement is low coding complexity, which means that the computational complexities of both encoding and decoding must be low. The third requirement is to be *stream-decodable* in the sense that each codeword symbol should be decoded with a low average latency, typically measured by the number of future codeword symbols that must be received to decode the current one. A code that is stream-decodable is called a *streaming code* in the literature [4, 6, 16, 34, 35, 46].

---

Authors' Contact Information: Qianru Yu, Georgia Tech, Atlanta, USA, qyu87@gatech.edu; Tianji Yang, Georgia Tech, Atlanta, USA, tyang425@gatech.edu; Jingfan Meng, Georgia Tech, Atlanta, USA, jeffmeng@live.com; Jun (Jim) Xu, Georgia Tech, Atlanta, USA, jx@cc.gatech.edu.

### 1.1 Motivations for Efficient Streaming Code

While the first two requirements are universally needed by today's high-speed networks and the applications running on them, the third requirement is essential for mission-critical and time-sensitive applications such as real-time video streaming, cloud gaming, remote telesurgery [39], and autonomous vehicle systems [23], where retransmission-based recovery is infeasible due to strict delay budgets (often below 100 ms). For example, there has been an increasing demand for truly real-time (TRT) one-to-many video streaming services [1, 36], such as live-broadcasting a soccer game. TRT video streaming demands an extremely low network latency that can be characterized, using this soccer game example, as follows: if you hear the cheering of your neighbor before you see a goal being scored, then your TV provider has failed you. In addition, there are also feedback-constrained scenarios, such as underwater (with seconds-long RTTs over kilometer distances) [24] and satellite links, in which retransmissions are costly or impractical.

Even for one-to-many streaming of a pre-recorded video, which has a much weaker latency requirement (tens of seconds) than TRT (truly real-time) as it allows generous use of buffering [41], ECC is the only viable packet loss recovery mechanism, as it is impossible for the streaming server to handle the (TCP) retransmission requests from many receivers [30, 31]. Luby transform (LT) code [30] is the state-of-the-art ECC solution for one-to-many file distribution and pre-recorded video streaming. However, existing LT codes (including an LT refinement called Raptor [49, 50]) are not well-suited for TRT video streaming because, to achieve very low latency required by TRT, a small *block size* (say, no more than a few hundred packets) has to be used, which results in high coding overhead (for reasons to be explained shortly).

Most existing ECC schemes are *block codes* in the following sense: each codeword consists of a block of codeword symbols, and both encoding and decoding operate on entire blocks, meaning that decoding *usually requires receiving all codeword symbols in a block*. Hence, when such a block code is used (at the packet level), the larger the block size (in number of packets), the longer the decoding latency, as the receiver need to collect all coded packets in a block. Existing work on streaming codes has relied almost exclusively on a block code called Reed-Solomon (RS) [42], since RS codes are Maximum Distance Separable (MDS) [52] and therefore optimal in terms of coding efficiency (given any fixed block size). However, achieving low decoding latency, a key requirement for streaming code, requires small block sizes, as explained above. This, in turn, reduces coding efficiency, a well-known trade-off in coding theory that can be illustrated by an analogy from the insurance industry: the smaller the risk pool (block size), the higher the insurance premium (coding overhead). Furthermore, RS codes are computationally expensive to encode and decode, as we will elaborate in Section 3.3.

This latency-efficiency trade-off also applies to non-block codes such as the aforementioned LT code [30], for the following reason. In coding theory, when an ECC scheme is described as "non-block", it usually means that the scheme does not have a fixed "$(n, k)$ design", where $n$ is the total number of codeword symbols (data plus redundancy), and $k$ is the number of source symbols (the original information block to be sent). More precisely, such an ECC scheme still has a certain $k$, but does not have a fixed $n$. However, this $k$ roughly corresponds to the *effective block size* as far as decoding latency is concerned. For example, in the case of LT codes, the receiver would have to receive close to $k$ codeword symbols before a nontrivial percentage of the $k$ source symbols can be recovered – a behavior that we refer to as the "zero-coupon bond" in Section 3.2.1. For this reason, when (hypothetically) using LT codes to stream a movie (which can contain millions of packets [7]), the content must be cut into many blocks (for separate encoding and decoding) so that the size $k$ of each block is not large enough to cause excessive decoding latency [30].

A dream solution to this open problem (in coding theory) would be a *native streaming code*, defined as one that allows each codeword symbol to be decoded on-the-fly (i.e., almost immediately upon arrival). As discussed above, existing streaming codes are not truly native; they are merely existing block codes (e.g., RS) operating with a small block size. This long-sought goal is finally realized by METTLE, our proposed solution introduced next.

## 1.2 METTLE: Multi-Edge Type with Touch-less Leading Edge

In this work, we propose METTLE, the first ECC scheme that simultaneously satisfies all three requirements. It naturally achieves the second requirement – low coding complexity – since it belongs to the same *peeling* family as Tornado [31] and LT codes [30]. Specifically, in METTLE, encoding or decoding each (coded) packet involves only a small, constant number (typically between 3 and 5) of peeling operations (bitwise-XORs between packet pairs), which matches the (low computation cost) record achieved by Tornado codes. In comparison, the aforementioned RS codes used in existing streaming-code designs have encoding and decoding times several orders of magnitude higher, as we will elaborate in Section 3.3.

Also similar to LT, METTLE achieves very high coding efficiency (first requirement). However, unlike LT, METTLE can do so without compromising stream decodability (third requirement). For example, using the same coding overhead, METTLE has the same erasure correction capability as an LT code with a large block size $4 \times 10^4$ (packets), yet its average decoding latency is only around 57 packets under 1% (coded) packet loss, as we will show in Section 4.2.4.

METTLE is the first native streaming code: it is not a block code with an artificially small block size, but instead enables genuine on-the-fly decoding. How METTLE manages to achieve this "native" status is one of its key innovations: the decoupling of block size (or its equivalence) from decoding latency. As discussed earlier, even non-block codes such as LT possess an *effective block size*, which is analogous to the size of the risk pool in the insurance analogy above. In METTLE, this effective block size is no longer coupled with decoding latency – a property rarely achieved by existing codes (block codes or else). In fact, METTLE's effective block size is the entire dataset to be transmitted (e.g., an entire movie file), yet its decoding latency remains extremely low and independent of that size. Thanks to such a humongous block ("risk pool") size, METTLE can achieve very high coding efficiency ("low insurance premium").

*1.2.1 Baseline METTLE: from Spatial Coupling to Time Coupling.* METTLE builds on Walzer's scheme [56], which belongs to the same *peeling family* and hence naturally achieves low coding complexities (second requirement above). Walzer's scheme [56] can be viewed as an improved invertible Bloom lookup table (IBLT) [22], with the sole intended improvement (as stated in [56]) being the coding efficiency, or greater "compactness," of the IBLT data structure. To achieve this improvement, Walzer's scheme essentially augments IBLT with a celebrated technique from coding theory called *spatial coupling* [19, 27–29].

After reading Walzer's paper last year, we were struck by the possibility that his scheme might, with sufficient coding overhead, be adapted into a streaming code for our applications. Exploring and discovering this connection is our contribution, as this connection is far from obvious in two respects. First, stream decodability, whether achievable or not, is irrelevant to Walzer's intended applications. Second, for his target applications – which require *blind encoding* capability (to be described in Section 2.1) – stream decodability is anyway impossible.

Walzer's scheme augments IBLT with spatial coupling as follows: (the equivalence of) each source symbol is mapped (by hashing) to a spatial location, say $\tau$, that corresponds to an index in the IBLT array, and is then encoded into a few codeword symbols uniformly distributed near $\tau$, or in other words, *spatially coupled* to a neighborhood of $\tau$. METTLE simply adapts *spatial coupling* in

Walzer's scheme to *time coupling* as follows. METTLE treats the first packet (e.g., in a movie file) as a source symbol "hashed to" spatial location $\tau = 0$, the second packet as that "hashed to" $\tau = 1$, and so on; and then encoding these packets in the same way as Walzer's scheme would do to source symbols that are hashed to these spatial locations. When a considerable amount of coding overhead (e.g., 21% overhead for reliably protecting against 1% packet loss) is used, such an adaptation (by METTLE) allows the time-ordered source symbols to be decoded from "left" (earlier in time) to "right" (later in time) in a largely single-pass manner.

*1.2.2 Adding MET and TLE to the Baseline.* METTLE adds two enhancements that can significantly reduce this coding overhead (e.g., from 21% to 5.5% against 1% packet loss in the example above): (1) MET (Multi-Edge Type) spatial coupling, and (2) TLE (Touch-less Leading Edge). Their acronym concatenation – METTLE – captures the spirit of our design: as we will explain shortly, METTLE "copes well with difficulties" (e.g., burst erasures) "in a resilient way," borrowing the phrasing from the dictionary definition of the English word *mettle*.

*1.2.3 Attractive properties of METTLE.* Beyond solving the open problem in coding theory, METTLE offers several other attractive properties. In the interest of space, we highlight three of them below.

*Hashing-based code.* Like Walzer's scheme, METTLE is a hashing-based code. That is, METTLE's code structure, or more precisely its underlying Tanner graph [43] (see Section 2.3), is defined implicitly via hashing (using a random seed) and generated on-the-fly during encoding (at the sender) and decoding (at the receiver). METTLE derives the following three important "systems benefits" from being a hashing-based code. First, METTLE's coding operation does not require storage of the code structure (in memory), making it nearly stateless and lighweight to implement. Second, METTLE is a *continuous rate* code in the sense that it can be configured to use any desired coding overhead ratio, unlike $(n, k)$ block codes whose overhead ratio is fixed to $(n - k)/k$ by design. Third, METTLE can vary this overhead ratio value on-the-fly in response to changing channel conditions (which is roughly equivalent to rate adaptation in a wireless network), without incurring much (network) protocol processing or systems configuration overheads. In a sense, METTLE allows seamless, continuous rate adaptation.

*Impressive mettle (resilience).* METTLE is also highly resilient to packet losses (erasures) under time-varying rates, of which burst erasures are a special case – conditions far harder to protect against than constant erasure rates. Specifically, when the erasure rate fluctuates within a wide range $[r_{\min}, r_{\max}]$ with an average (rate) $r_{\text{avg}}$, METTLE can effectively protect against such variation using only slightly higher coding overhead ratio than it would need for the constant erasure rate $r_{\text{avg}}$. In contrast, in this case, a block code such as Reed-Solomon [42] would require either continuous estimation of the time-varying erasure rate [3, 47], a difficult and error-prone task, or provisioning against (near-worst-case) erasures at constant rate $r_{\max}$, which significantly reduces the coding efficiency. This ability to "budget for the average erasure rate" makes METTLE especially compelling for wireless network applications such as real-time video streaming (to smartphones), where packet losses (after CRC error) can occur at widely varying rates (say due to fast fading [53]), even with strong bit-level coding (using LDPC or Turbo) per packet and prompt rate adaptation.

A natural question is what happens if the actual long-term average erasure rate happens to exceed the "budgeted rate" of the application using METTLE. In this case, decoding may eventually stall at a certain point, prompting receiver feedback (if the return channel exists). The sender can then increase the coding overhead ratio to match the higher average erasure rate, using METTLE's seamless, continuous rate adaptation described above. Alternatively, applications can budget conservatively for a much higher average rate at a modest cost, thanks to METTLE's high

coding efficiency. For example, to reliably protect against 2% loss, METTLE requires 7% coding overhead, which is only slightly larger than the 5.5% overhead needed for 1% loss.

*Systematic METTLE.* Finally, consider the harsher case where (1) the actual average erasure $r_{\text{bad}}$ still exceeds the conservatively "budgeted rate," and (2) no return channel exists (e.g., one-to-many live streaming or deep-space communications). METTLE has one last line of defense: thanks to TLE, any METTLE codeword can be written in systematic form (i.e., containing a verbatim copy of the original data) without sacrificing coding efficiency, a property not shared by all ECC schemes. With systematic METTLE, the receiver may fail to recover at most $r_{\text{bad}}$ (percent) of the source data, which remains tolerable in many applications (e.g., watching videos on a smartphone).

## 2 METTLE Coding Scheme

In this section, we describe the design of the METTLE scheme. We present it before delving into the coding theory background, since only minimal theory is needed to grasp *what* METTLE is, *how* it works, and *why* it is designed that way. The necessary background will instead be introduced in Section 3, where we position METTLE's contributions within the broader tapestry of the coding theory literature.

As mentioned earlier, METTLE builds on Walzer's scheme [56], which itself extends the IBLT (Invertible Bloom Lookup Table) [22]. Accordingly, our presentation follows the path IBLT (Section 2.1)→Walzer (Section 2.2)→METTLE (Section 2.3). To make the exposition clearer, we describe all three schemes using the balls-into-bins language.

### 2.1 IBLT

We present only the vanilla IBLT that Walzer's scheme extends, as none of its more sophisticated variants is directly related to this work [25, 32, 45]. Similar to Bloom filter (BF) [8], an IBLT also encodes a set $\mathcal{S}$, but can do much more than a BF (on $\mathcal{S}$): Whereas a BF can answer whether a given (query) element $x$ belongs to $\mathcal{S}$, an adequately-sized IBLT can be decoded to enumerate all elements contained in $\mathcal{S}$. IBLT's data structure, encoding procedure, decoding procedure, and primary application are as follows.

*2.1.1 Data structure.* An IBLT data structure encoding a set $\mathcal{S}$ is simply an array of $(1 + c)n$ bins (called buckets in the literature), where $c > 0$ corresponds roughly to its (coding) overhead ratio and $n \approx |\mathcal{S}|$. Each bin contains three fields: (1) an identifier, (2) a counter, and (3) a signature. To avoid dealing with messy notations, we avoid using floors or ceilings (e.g., over $(1 + c)n$ here) throughout Section 2, and ask readers not to worry about our using a possibly non-integer (due to not applying floors or ceilings to it) where an integer is called for (e.g., the array index of a bin).

*2.1.2 Encoding procedure.* Each element in $\mathcal{S}$ is considered a ball. The encoding process throws all these $|\mathcal{S}|$ balls, each ball being thrown a constant number (say $l$) of times, into these bins, as follows. Each ball is thrown into $l$ bins whose indices are i.i.d. random variables (RVs) uniformly distributed in the IBLT's (array) index range $[0, (1+c)n)$ determined by $l$ different hashing functions $h_i$, $i = 1, \cdots, l$. After all the $|\mathcal{S}|$ balls are thrown into the IBLT, its state is the values of its $(1 + c)n$ bins. The identifier (field value) of each bin is the bitwise-XOR of the identifiers of all balls hashed into it.

*2.1.3 Decoding procedure.* The decoding procedure is iterative. In each iteration, it locates all bins containing exactly one ball (inferred through the counter field). Each such ball is thus "exposed" (with its identifier say $x$ revealed) and removed (peeled) from all $l$ bins (calculated by applying the hash functions to $x$, and verified by the signature field) that it is thrown into. The peeling operations in the current iteration may cause more balls to be exposed and peeled in the next iteration, which

propels the peeling process. The decoding is successful if every bin become empty (contains no balls) when peeling terminates. Intuitively, the coding overhead ratio $c$ needs to be large enough for successful decoding with high probability. Indeed, with the optimal parameter $l = 3$, successful peeling requires $c \geq 0.222$ [22], when $n$ (and correspondingly $|\mathcal{S}|$) is sufficiently large.

*2.1.4 Primary application.* IBLT was originally proposed to solve the set conciliation problem [15, 18, 57], which remains its primary application. In this problem, two large sets A and B of objects (tokens, files, records, etc.) are stored respectively at two different network-connected hosts, which we name Alice and Bob respectively. Alice and Bob need to communicate with each other to learn the set union $A \bigcup B$ (by learning the set difference), and to do so at low communication and computation costs. An IBLT-based set reconciliation solution works as follows. Alice first sends its IBLT(A) to Bob, who then "subtracts" its IBLT(B) (configured with the same $n$, $c$ and hash functions) from it. It can be shown that the resulting IBLT difference encodes precisely the set difference $|A \triangle B| \triangleq (A \setminus B) \bigcup (B \setminus A)$, which Bob can recover through the peeling process if the IBLT is properly sized (in accordance with $|A \triangle B|$).

We explain a subtle but crucial fact that lies at the heart of METTLE's TLE innovation. By independently computing their IBLTs and then letting Bob decode the IBLT difference, Alice and Bob collaboratively performs a *blind encoding* of the set difference: the encoders (Alice and Bob) encode a set $A \triangle B$ without either knowing the full list of elements. As we will explain in Section 2.1, TLE cannot be used by Walzer's scheme for Walzer's applications (to reduce the coding overhead), since it is not compatible with the *blind decoding* requirement. We take credit for observing that our applications do not require blind decoding and for inventing TLE to fully exploit the dispensability of this requirement in our applications.

## 2.2 Walzer's Scheme

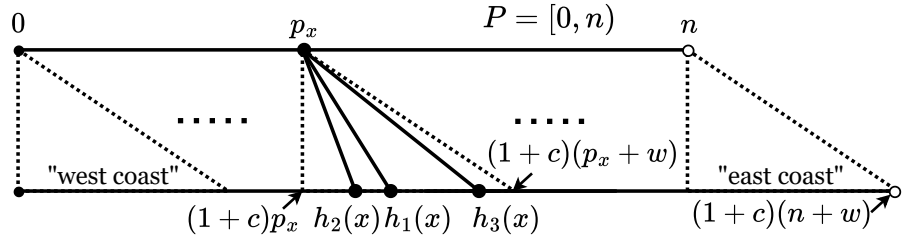

Fig. 1. Walzer's Scheme. For each position $p_x \in P = [0, n)$, select $l = 3$ incident nodes (bins) independently and uniformly at random from the range $[(1+c)p_x, (1+c)(p_x + w))$. During peeling decoding, the "west" and "east" coasts erode inward from both ends at each iteration. In other words, the decoding front advances from both sparse-degree ends toward "inland", peeling the unrecovered region layer by layer.

Walzer's scheme enhances IBLT by reducing its overhead ratio $c$, thereby improving coding efficiency. It achieves this through a celebrated technique from LPDC coding theory known as spatial coupling [19, 27–29], as follows. Again, like the IBLT, the data structure encodes approximately $n$ balls, but now consists of $(1+c)(n+w)$ bins, where $w > 0$ denotes the (spatial-) *coupling window* size. In (Walzer's) theoretical model, $n$ is assumed to approach $\infty$ and $w = o(n)$, so the overhead ratio remains to be $c$ in the limit.

A key difference between Walzer's scheme and the IBLT lies in the range of the $l$ (independent) uniform hash functions $h_i(\cdot)$, $1 \leq i \leq l$. In the IBLT, each $h_i$ maps uniformly over the entire index range $[0, (1+c)n)$, meaning any ball $x$ can land anywhere in the data structure. In contrast, Walzer's

scheme localizes each $h_i(x)$ to a small neighborhood of width $(1+c)w$ in its index range $[0, n)$, as follows. This ball $x$ is first mapped to a uniform random *position* $p_x$ in the "ball range" $[0, n)$ using a different uniform hash function than $h_j(\cdot), 1 \le j \le l$. Then, for each $i$, the hash value $h_i(x)$ is uniformly distributed in the range $[(1+c)p_x, (1+c)(p_x+w))$, or in other words *coupled* to the right *spatial* neighborhood of $(1+c)p_x$.

This *spatially coupled* hashing in Walzer's scheme is illustrated in Fig. 1. The top line represents the "ball range" $[0, n)$, and the bottom line represents the "bin range" $[0, (1+c)(n+w))$. To visualize the concept of spatial coupling, the bottom line is scaled by a factor of $1/(1+c)$, so that a ball at position $p_x$ on the top line aligns vertically with the bin at position $(1+c)p_x$ on the bottom line.

*For ease of presentation, throughout the rest of Section 2, we assume $l = 3$ hash functions $h_1(\cdot), h_2(\cdot), h_3(\cdot)$ are used, unless stated otherwise. In practice, both Walzer's scheme and METTLE typically use between 3 and 5 hash functions.* As shown in Fig. 1, the three hash values $h_1(x), h_2(x), h_3(x)$ are uniformly distributed in the range of $[(1+c)p_x, (1+c)(p_x+w))$ on the bottom line. Fig. 1 also shows the hash range of a hypothetical ball near the position $n$ on the top line, which explains why Walzer's scheme requires an additional $(1+c)w$ bins compared to the IBLT.

Similar to IBLT, Walzer's scheme also uses the iterative peeling decoder (referred to as *parallel peeling* in [56]). However, with the same overhead ratio $c$, Walzer's scheme is "easier to peel" than the IBLT due to a "nucleation effect" introduced by spatial coupling. Consider the leftmost bin (at position 0): the number of balls falling into this bin follows a Poisson distribution with mean $\frac{3}{(1+c)w}$ (assuming $l = 3$), which is close to 0 when $w$ is not small. This mean (value) then increases linearly and gradually until it reaches $\frac{3}{1+c}$ at bin position $(1+c)w$. As such, bins in the "west coast" region $[0, (1+c)w)$ are more likely to contain exactly one ball (and hence can be decoded) than bins "inland." Peeling those bins removes their corresponding balls from other, more densely populated bins (that are further "inland"), triggering the "erosion of the west coast," and exposing a new eady-to-peel "west coast." In [26, 27], a chain reaction similar to what we see here is likened to the freezing process of the super-cooled (to below 0 °C) pure water in a smooth container: it remains liquid until a nucleus (e.g., dirt) initiates crystallization, after which freezing propagates rapidly. In Walzer's scheme, the "thinly populated bins" (by balls) on the "west coast" serve the role of such nuclei. In fact, the "east coast" region $[(1+c)n, (1+c)(n+w))$ is similarly "thinly populated", as readers may have noticed, and also serves the same role. Hence, Walzer's scheme can (easily) be peeled from "both ends." Due to the "nucleation effect" brought by spatial coupling, Walzer's scheme achieves successful peeling with a smaller overhead ratio $c$ than the IBLT [56].

### 2.3 METTLE

As mentioned above, Walzer's scheme was proposed solely to improve the coding efficiency of hashing-based information retrieval data structures such as IBLT. In Walzer's paper [56], there is no indication that streaming codes were considered as an application or even as a possible direction. Moreover, the parameter choices inherent to Walzer's design—for example, how the coupling width $w$ scales with the "block size" $n$ to optimize the overhead ratio $c$—fundamentally preclude its adaptation into a streaming code, as we will elaborate shortly.

As mentioned earlier (in Section 1.2.1), METTLE adapts Walzer's scheme into a streaming code, through three major modifications: (1) from spatial coupling to "time coupling," (2) TLE (touch-less leading edge), and (3) MET (multi-edge type). We describe them in Sections 2.3.1 through 2.3.3. Finally, Section 2.3.4 introduces the systematic METTLE variant.

*2.3.1 From spatial coupling to "time coupling".* As mentioned earlier, METTLE inherits Walzer's spatially coupled IBLT data structure, but converts its space coupling to "time coupling." Consider a block of $n$ time-ordered packets where $n$ can be arbitrarily large and may not be known in advance.

In METTLE, the encoding of these packets works as follows. Upon the arrival of the first packet (ball), METTLE assigns to it the (spatial) ball position 0 (on the top line in Fig. 1), and accordingly throws it into 3 (we have fixed the number of hash functions to $l = 3$ as "announced" in Section 2.2) bins independently and uniformly distributed in the bin range $[0, (1+c)w)$ (that is spatially coupled to its ball position 0). Similarly, the second packet is placed at ball position 1 and thrown into 3 uniform random bins in the bin range $[c, (1+c)(1+w))$ (that is spatially coupled to its ball position 1), and so on. In other words, the position of a ball is no longer determined by hashing (using a different hash function than $h$) as in Walzer's scheme but by the time order (sequence number) of the ball. Hence, we refer to $w$ as the *time-coupling window*. METTLE can place these $n$ packets precisely at these $n$ ball positions since we know their "identities" and the exact (time) order among them, which we cannot do in Walzer's scheme due to the *blind encoding requirement* described in Section 2.1.

We now describe how METTLE is used in a network operation (between a sender and a receiver). At the sender's side, the coded packets (contents of the bins) are released and transmitted to the receiver, at a rate that is $(1 + c)$ times faster than the (uncoded) packets arrived at the sender (say from the upper layer in the network stack of the sender). That is, right after the arrival of the $t^{\text{th}}$ (uncoded) packet, a cumulative total of $(1 + c)t$ coded packets can be released to the receiver (unless the sender's computer system becomes a performance bottleneck). This smooth operation ($(1 + c)t$ "coded departures" after $t$ "uncoded arrivals," for any $t$), in combination with the TLE technique ensuring that the $t^{\text{th}}$ "uncoded arrival" is encoded into the $((1 + c)t)^{\text{th}}$ "coded departure" (to be explained shortly) and hence can be "immediately" decoded by the receiver (under lossless channel), allows METTLE to achieve zero *latency under lossless channel* (defined as no extra delay imposed artificially by the streaming code [46]).

At the receiver's end, the receiver performs on-the-fly decoding when the coded packets stream in, as follows. The receiver tries to peel the set of received coded packets, mostly from left ("earlier in time") to right ("later in time"), except when the loss of an earlier coded packet can be recovered backward using later-arriving coded packets. To facilitate this peeling operation, each coded packet carries a *sequence number* indicating its corresponding bin index.

Now we elaborate on METTLE being a hashing-based code (with three major benefits) as claimed in Section 1.2.3. As shown in Fig. 1, METTLE can be viewed as–and indeed is–a random bipartite graph code, with the balls being one vertex set, the bins being the other vertex set, and the "(which) balls (are thrown) into (which) bins" relationships being the set of edges between the two vertex sets. This bipartite graph can be called a Tanner graph [54] (although Tanner graph is a term mostly reserved for LDPC codes [21], whereas METTLE arguably belongs to the LDGM (low-density generator matrix) [43] code family). As just explained, this Tanner graph (the edges therein) is computed on-the-fly, by hashing on (the sequence numbers of) the original (uncoded) packets.

*Encoding and decoding complexities.* To encode a packet (ball) involves only $l$ hashing operations and $l$ packet-wise XOR computations (for throwing the ball into $l$ random bins determined by hashing), where $l$ is typically a small number between 3 and 5 as explained above. To decode a ball, and remove it from other bins, involves $l$ hashing operations and $l$ packet-wise XOR computations on average. As such, METTLE's coding complexities are extremely low.

*Coding efficiency.* This baseline METTLE, which is essentially Walzer's scheme with one-directional (rightward) peeling, is acceptable in terms of coding efficiency, as the following experimental results show. At a (coded-packet) loss rate of 0, all original (uncoded) packets can be recovered (from decoding the received coded packets), with probability at least 0.999, when the overhead ratio $c$ exceeds the value 0.2. The value $c = 0.2$ is only slightly larger than $c = 0.17$ needed by Walzer's scheme (with bi-directional peeling) to achieve the same success probability, since the base METTLE

can peel only from left to right. As the loss rate increases to 1%, the $c$ value increases to 0.21 for the baseline METTLE.

Note also that the $c$ values needed by both the baseline METTLE and Walzer's scheme are much larger than the theoretical limit $c = 0.089$ reported in [56]. There are three reasons to this. First, this theoretical limit does not "budget for" a high success probability (like 0.999 here). Second, in Walzer's scheme, for optimizing $c$ (toward this theoretical limit), the coupling window size $w$ scales as $\Theta(n^{2/3})$. METTLE cannot use such a large $w$ (and hence its $c$ "suffers" as a result) when $n$ is large, since the decoding latency grows roughly linearly with $w$. Third, this theoretical limit can be approached only when $n$ is humongous.

*2.3.2 Add MET to the baseline.* While the baseline METTLE provides a good starting point, requiring a coding overhead of $c = 0.21$ to handle a 1% packet loss may seem excessive. We propose two techniques, namely MET (multi-edge type) and TLE (touch-less leading edge), that serve as lubricants to facilitate smoother (left-to-right) peeling. With these two enhancements, the overhead ratio can be reduced to around 5.5% under 1% packet loss rate, and the average decoding latency is only a few tens of packets.

Before we describe these two enhancements, we simplify the notations by dropping the (notation for the) hashing functions in this subsection. Recall that in the baseline METTLE, a packet (ball) at the ball position $x$ (corresponding to the $(x+1)^{\text{th}}$ packet) is thrown into three bins whose indices $h_1(x), h_2(x)$, and $h_3(x)$ are RVs that are independently and uniformly distributed over the bin range $[(1+c)x, (1+c)(x+w))$ spatially coupled with (the ball position) $x$. The upcoming MET enhancement refines this design by tuning the distributions of these RVs to optimize the coding efficiency of METTLE. Since, at this stage, we are no longer concerned with how these RVs are generated (i.e., through hashing), we replace each explicit hash function notation $h_i(x)$ with a variable $\eta_i$ that captures the same information.

Fig. 2 zooms into the window shown in Fig. 1, using a stretched horizontal scale to better illustrate the relative edge offsets to the right window boundary and the new notation $\eta$. These three RVs, which encodes (the positions of) the three bins that the ball $x$ is thrown into, are denoted as $\eta_1, \eta_2$, and $\eta_3$ respectively. Each $\eta_i$, $i = 1, 2, 3$, is the distance of the $i^{\text{th}}$ bin position from $(1+c)(x+w)$, the right boundary of the time-coupling window. In the baseline METTLE, these three RVs are i.i.d. uniform in the range $[(1+c)x, (1+c)(x+w))$.

Through extensive experiments, we observed that making these three (mutually independent) RVs not identically distributed can improve the coding efficiency slightly. Fig. 2 shows a realization of these three RVs according to a near-optimal set of distributions we had identified before discovering

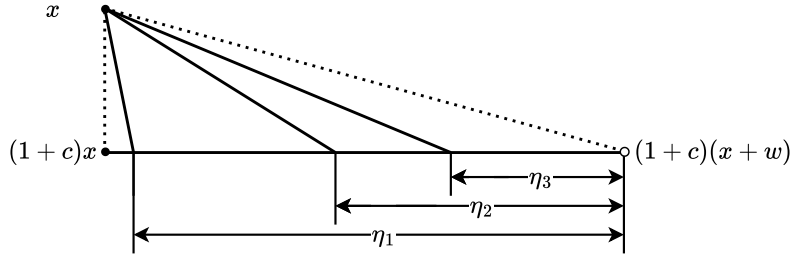

Fig. 2. Illustration of MET coupling: a source symbol at $x$ connects to three landing positions within its time-coupling window $[(1+c)x, (1+c)(x+w))$, with offsets $\eta_1, \eta_2, \eta_3$ measured from the right boundary. For clarity, the coupling window is stretched horizontally and not drawn to scale.

the TLE technique. Specifically, the probability mass of $\eta_1$ is concentrated near (but not exceeding) $(1 + c)w$; $\eta_2$ follows Binomial$((1 + c)w, 1/2)$ distribution; and $\eta_3$ follows Binomial$((1 + c)w, 1/4)$. As such, the expectations of $\eta_1, \eta_2, \eta_3$ decrease roughly (with the "roughness" coming from $\eta_1$) exponentially, or in other words the corresponding bin indices become roughly exponentially closer to $(1 + c)(x + w)$, the right boundary of the time-coupling window.

In retrospect, such an "exponential decay" in distance (toward the right boundary) should improve coding efficiency compared to the i.i.d. uniform used in the baseline METTLE, for the following intuitive reason. This "exponential decay" makes the "west coast" more "sparsely populated" (bins containing less balls statistically) and hence easier to peel from left ("west") to right ("east"), than making them i.i.d. uniform. However, to our disappointment, the improvement in coding efficiency using this "exponential decay" is small: the overhead ratio $c$ decreases slightly from 0.21 to 0.18 (for a success probability of 0.999) under 1% loss. Motivated by this underwhelming result, we kept looking and eventually discovered the TLE technique (to be described next in Section 2.3.3) that can make a huge difference.

In coding theory, letting $\eta_1$, $\eta_2$, and $\eta_3$ (the relative positions of the vertices on the bottom line that incident to the edges fanning out from the ball $x$ on the top line) to follow different distributions can be viewed as a degenerate form of the Multi-Edge Type (MET) framework. We actually have no choice but to call this enhancement MET for a mathematical reason: its density-evolution (to be introduced in Section 3.4) formula is a vector-recursion equation characteristic of a MET-LDGM code, of which METTLE is arguably an instance.

*2.3.3 Touch-less Leading Edge (TLE), the final touch.* As shown in Fig. 2, $\eta_1$ is close to the left boundary of the time-coupling window. The TLE technique is to simply let $\eta_1 = (1 + c)w$; as such, the RV $\eta_1$ becomes a constant. In this case, the "first edge" becomes deterministic and is vertically incident on the bin located at the left boundary of the time-coupling window. Similarly, for the ball at each position $y$ ($y = 0, 1, 2, \cdots, n - 1$), we define $h_1(y) = (1 + c)y$ (now we "need the first hash function back"), meaning that the "first edge" fanning out from the ball at position $y$ is always incident on the bin $(1 + c)y$.

This "TLE hash function" $h_1(\cdot)$ has two important properties. First, the mapping is one-to-one, or the "first edges" fanning out from any two distinct balls will be incident on different bins, because two neighboring incident positions (bins) have distance $1 + c > 1$ between them. We say these "first edges" are touch-less because no two of them will touch each other (fall into the same bin). Second, right after the arrival of a ball $y$, the coded packet at the bin $(1 + c)y$, which is incident by $y$'s "first edge," can be released right away (to the receiver), since $y$ is provably the last packet to be XOR-ed with the (content of the) bin. As explained in Section 2.3.1, this "immediate release" property allows METTLE to have zero latency under lossless channel. We name the "first edge" the leading edge, since it leads in time compared to other edges fanning out from the same ball.

Adding TLE to the baseline METTLE (which does not have MET) improves its coding efficiency significantly. For example, as we will illustrate in Section 4.2.1, the overhead ratio $c$ can be improved from 0.21 to 0.07 under 1% packet loss rate.

Readers may wonder why we do not also make $\eta_2$ and $\eta_3$ constants (e.g., $\eta_2 = (1 + c)w/2$ and $\eta_3 = (1 + c)w/4$). We have in fact tried this and found that the coding efficiency drops significantly when $\eta_2$ and $\eta_3$ are not randomized. We now know the reason from the coding theory literature: the resulting Tanner graph exhibits a poor girth structure (i.e., contains many short cycles), which in turn portends poor coding efficiency. Interestingly, making $\eta_1$ deterministic in the TLE manner while randomizing $\eta_2$ and $\eta_3$ does not cause this girth problem, as we will show empirically in Section 4.2.5. An intuitive explanation is that it takes more than one "deterministic edge" per ball to induce numerous short cycles that will reduce the coding efficiency.

*2.3.4 Systematic Variant.* METTLE as described above is clearly not a systematic code. However, it can always be converted to a systematic form at no cost in coding efficiency and complexity, which comes as a side benefit of TLE. The conversion works as follows. Let $p_1, p_2, \cdots, p_n$ be the actual contents of $n$ time-ordered balls that, are placed at ball positions $0, 1, \cdots, n-1$, respectively. As just explained, the TLE (edges) of these $n$ balls are incident on the bins $0, 1+c, 2(1+c), \cdots, (n-1)(1+c)$ respectively. We call these $n$ bins the *TLE bins*.

We make two closely related claims, which together express essentially a single idea. First, we can convert this coding instance into the following systematic instance: the original ball-contents $p_1, p_2, \cdots, p_n$ appear in the $n$ TLE bins in their original sequence (time) order – that is, $p_j$ appears in the $j^{th}$ TLE bin $(j-1)(1+c)$. Second, this systematic instance can be obtained by (encoder) placing "fake ball contents" $q_1, q_2, \cdots, q_n$ at the $n$ ball positions. These two claims follow immediately from the following fact that we will prove shortly: the $n \times n$ 0-1 matrix associated with the $n$ XOR-linear balls-into-bins equations relating $p_1, p_2, \cdots, p_n$ (in the $n$ TLE bins) to $q_1, q_2, \cdots, q_n$ (in the $n$ ball positions) is full-rank and hence "solvable." Once $q_1, q_2, \cdots, q_n$ are thus "solved," the contents of $(1+c)w + cn$ other (non-TLE) bins (which serve as repair packets for the "systematic (coded) packets" $p_1, p_2, \cdots, p_n$) are determined accordingly.

Now we explain what this matrix is, from which this fact follows immediately. This matrix is actually the generator matrix of METTLE restricted to the $n$ TLE bins, which we denote as $\mathcal{G}$. Every diagonal element of $\mathcal{G}$ has value 1, which corresponds to the one-to-one correspondence between the $n$ balls and the $n$ TLE bins. $\mathcal{G}$ is upper-triangular because, as explained in the second paragraph in Section 2.3.3 (the second property of $h_1(x)$), a TLE (edge) is the last edge (in time) to hit a bin (so "nothing can go under" in $\mathcal{G}$). Hence, $\mathcal{G}$ is full-rank.

## 3 Background and Related Work

This section surveys background and related work, focusing on results most pertinent to METTLE. Since METTLE is an erasure code–designed to protect against packet losses (erasures)–we primarily situate the discussion within the erasure-channel model.

### 3.1 Binary Erasure Channel (BEC) and its variant

A memoryless packet-erasure channel erases each packet (symbol) independently with a constant probability $\varepsilon$. The term *memoryless* refers to the erasure events being independent. This model is analogous to the well-known binary erasure channel (BEC), except that the erased unit is a packet rather than a bit. Hence, with slight abuse of terminology, we refer to this model as BEC, and denote it as $BEC(\varepsilon)$ when convenient.

As mentioned earlier, erasures can also occur at a time-varying rate with its extreme form being the burst erasures (consecutive packet losses), and METTLE is particularly resilient to such more challenging erasure patterns. Such erasures will be modeled as a Gilbert-Elliott (GE) [17] channel, to be described in Section 4.1.2, for evaluating METTLE.

### 3.2 Luby transform (LT) and Raptor codes

*3.2.1 LT codes.* Luby transform (LT) code [30] is the first practical realization of *fountain codes* [10]. Like METTLE, LT code [30] also belongs to the decoding-by-peeling family, and hence is computationally lightweight for both encoding and decoding. Each codeword symbol (bin) is the XOR of a random subset of source symbols (balls), where the subset size (degree) follows the *robust Soliton degree distribution*. However, the average degree (number of balls in a bin) in LT is $O(\log n)$, where $n$ is the (information) block size (in number of source symbols), while in METTLE it is a

small constant no more than 5. Hence, LT codes have higher encoding and decoding complexity compared to METTLE.

LT codes are known for their rateless property—a set (block) of $n$ source symbols can generate, in principle, an unlimited stream of encoded symbols (when $n$ is not small). The receiver can recover the original data once slightly more than $n$ encoded symbols have been collected. However, LT achieves high coding efficiency only for the large block size $n$ (on the order of 10,000 symbols). For example, to match METTLE's coding efficiency, $n$ must be around 40,000. Such a large block size, however, leads to a high decoding latency of roughly 40,000 packets (meaning LT is not stream-decodable). This is because LT's decoding progress behaves like a zero-coupon bond: no "principal or interest payments" (i.e., few source symbols can be decoded) occur until "maturity" (when slightly fewer than $n$ coded symbols have been received). Sliding-window variants of LT codes [9, 11] reduce this decoding latency by only a factor of two to five, at the cost of increased decoding complexity.

*3.2.2 Raptor codes: Truncating the long tail of LT.* When the block size $n$ is small, LT codes suffer from low coding efficiency, primarily due to a phenomenon known as the "tail problem": decoding the final few source symbols can require a disproportionately long wait for additional coded symbols that happen to cover them. Raptor codes [49, 50] address this issue by truncating the tail, thus dramatically improving efficiency even at small $n$. Raptor codes achieve this by first encoding the $n$ source symbols into $n' > n$ intermediate symbols using a precode, which is a combination of a high-rate LDPC (Low-Density Parity-Check) code and a HDPC (High-Density Parity-Check) code. These intermediate symbols are then treated as the input to the LT encoding process.

In coding theory terms, Raptor codes are a serial concatenation of a structured inner code (LDPC+HDPC) and an outer LT code. This design allows the decoder to avoid recovering all $n'$ intermediate symbols via LT decoding. Instead, the inner precode is constructed so that the original $n$ source symbols can be recovered even if only about 98% of the intermediate symbols are obtained. Thus, the decoder can terminate early, avoiding the long tail of diminishing returns that plagues standard LT decoding at small block sizes.

Raptor pays a higher decoding complexity (cost) for the improved coding efficiency, for the following reason. To achieve efficiency beyond what pure peeling can offer, the LT decoding stage needs to resort to a process called *inactivation decoding* in Raptor specification [50]. Its inactivation decoder temporarily marks some symbols as inactive to let the peeling process continue; after peeling completes, the remaining inactive symbols are resolved using Gaussian elimination, adding a nontrivial computational cost. Moreover, decoding the HDPC part of the inner code also requires Gaussian elimination.

Raptor can, in principle, be configured as a streaming code, by using a small block size (e.g., $n$ = 200 packets). Hence, we compared its (the R10 variant's) coding efficiency with METTLE's, under a BEC with a constant loss rate of 0.1, targeting a decoding success probability of 99.9%. Both codes have similar latency (about 200 packets), yet METTLE requires a smaller coding overhead ($c = 0.256$) than Raptor ($c = 0.27$). This is a hard-earned win for METTLE, given that Raptor is allowed to use Gaussian elimination, while METTLE relies solely on peeling. METTLE's coding efficiency advantage can be explained by the "$3\sigma$ effect" (to be detailed next), which all block codes suffer from when the block size $n$ is small. Furthermore, this advantage becomes even more pronounced under highly variable loss rates that fluctuates widely, in which case Raptor must provision for near-worst-case conditions, whereas METTLE needs only to budget slightly above the average rate (as discussed earlier), a disparity further magnified by the $3\sigma$ effect. To summarize, METTLE is overall a better streaming code than Raptor because it has lower decoding complexity and higher coding efficiency, while achieving comparable decoding latency.

*3.2.3  3σ effect.* When the block size $n$ is small, random variation in the actual number of erasures within a block becomes significant. At a fixed packet loss rate $\varepsilon$, the number of erasures $X$ in a block of size $n$ follows a Binomial distribution: $X \sim \text{Binomial}(n, \varepsilon)$. Under the Gaussian approximation to the Binomial tail, the 99.9th percentile of $X$ can be estimated as $\mu + 3.09\sigma$ where $\mu = n\varepsilon$ is the mean and $\sigma = \sqrt{n\varepsilon(1-\varepsilon)}$ is the standard deviation. To achieve a decoding success probability of 99.9%, a block code must therefore be configured to correct up to at least $\mu + 3.09\sigma$ erasures. In other words, the required erasure correction capability must exceed the average by approximately $3\sigma$ effect in the sequel. Tooth similarly observed this phenomenon [3], which motivated Tooth to adjust the coding overhead on a per-codeword (per-frame) basis, according to the block size, rather than use a single conservative overhead ratio for all codewords.

## 3.3  Reed-Solomon and Streaming codes

Reed-Solomon (RS) codes [42] are often used as building blocks for streaming codes [4, 6, 16, 34, 35, 46]. When a small block size is used (e.g., $n$ =96 in [47]) to achieve stream decodability, RS achieves only slightly higher coding efficiency than METTLE (partly due to the $3\sigma$ effect above) under a constant erasure rate; and RS has lower coding efficiency under highly variable erasure rate, thanks to METTLE's ability to "budget for (slightly above) the average rate." Moreover, when operating at the packet level, which requires the column-wise implementation [13, 51], RS is two to three orders of magnitude more computationally expensive to encode and to decode than METTLE.

Streaming codes often use RS codes in a diagonal interleaving fashion [4, 6, 16, 34, 35, 46]. Specifically, the set of packets used to compute repair packets (which form part of an RS codeword) corresponds to a diagonal stripe that spans across parallel "pillars"–where each pillar represents a time-ordered frame of packets, and each packet within the pillar may be visualized as a "brick." The pillars themselves progress left to right in time, while the bricks are ordered top to bottom within each frame. As a result, the repair packets sent alongside each pillar belong to different RS codewords, each formed along a diagonal. This diagonal interleaving effectively mitigates burst erasures, because even a large burst that wipes out one or two pillars, each codeword only experiences erasures slightly and uniformly. Consequently, each source message (i.e., original set of data packets) can still be recovered by its decoding deadline [6, 20, 34, 35].

Recently, such RS-based streaming codes have been applied to latency-sensitive applications such as videoconferencing (Tambur [47]) and cloud gaming (Tooth [3]). However, the coding overheads reported in both Tambur and Tooth are relatively high—Tambur exhibits a median overhead of roughly 50%, while Tooth generally reports 10-30% overhead, even when operating in network environments with time-varying erasure rates that are, on average, significantly lower (e.g., largely between 1% and 10% as stated in Tambur [47]). In the case of Tooth, this high overhead can be attributed to two primary factors. First, as explained in Section 1.2.3 , Tooth uses per-frame RS codes (not a native streaming code), which requires the code parameters to be configured for the near-worst-case loss rate, rather than the average. This, in conjunction with the "$3\sigma$ effect" above (in Section 3.2.3), leads to substantial coding overhead. Second, in an attempt to improve efficiency, Tooth incorporates a predictive mechanism that estimates network dynamics—such as instantaneous loss rates and their moving averages. However, such prediction-based adaptation can be inherently error-prone, especially under volatile conditions, further degrading efficiency. In contrast, METTLE avoids both of these pitfalls: METTLE only needs to be configured for a slightly elevated erasure rate above the long-term average, and does not rely on any form of network prediction. This design simplicity contributes to METTLE's robustness and lower coding overhead, especially in environments with unpredictable or bursty losses.

### 3.4 LDPC, Spatial-Coupled LDPC, and Their Analysis

Low-density parity-check (LDPC) codes [21, 33] are (Shannon) capacity-approaching block codes. Under the BEC, the BP decoder of LDPC codes degenerates into a peeling decoder. Hence, LDPC has coding complexities comparable to those of METTLE when both operate at the packet level. This makes it particularly interesting to compare METTLE with large-block-size LDPC codes in terms of coding efficiency (Section 4.2.3). If METTLE does not lose much efficiency while achieving stream decodability, it truly earns a star—and indeed, it does.

As shown in Section 2.3, METTLE is also characterized by a Tanner graph [54], which was originally introduced in the context of LDPC. Hence, Tanner graph-theoretic techniques for analyzing LDPC code properties and performance, such as girth analysis, are also applicable to METTLE. We will perform a girth analysis on METTLE in Section 4.2.5, which shows that METTLE's hash-generated Tanner graph is pretty good.

The spatial coupling (SC) technique was invented originally for LDPC [12, 19, 28, 29, 37, 38, 48]. Mathematical machineries concerning SC, such as solving the density evolution (DE) [44] functional recursion equations using potential function theory (PFT), were also developed in the SC-LDPC theory literature [29, 59, 60]. Although similar in spirit, Walzer's SC takes a very different form than that in SC-LDPC, which may explain why Walzer (and coauthor) reinvented the idea in an earlier paper [14].

In [56], Walzer solved the DE recursion equations of his scheme using a similar PFT as used in SC-LDPC [29]. Hence, we attempted to apply the same approach to METTLE's DE recursion equations. However, this proves futile for the following reason. Since METTLE is a MET (mutli-edge type) code, its DE equations form a vector (functional) recursion rather than the scalar recursion in Walzer's SET (single-edge type) scheme: all $l$ hash values (edges) are i.i.d. uniform (within the spatial coupling window), and thus of the same type. There is unfortunately limited theory on the vector recursion form, originating from [60] and then arising from the DE analysis of SC-MET-LDPC [40]. The state-of-the-art framework for coupled vector recursions [40, 60] require the vector recursions to be symmetric and monotone, which METTLE's DE equations provably do not satisfy.

As part of our research on METTLE, we numerically "solved" these DE equations and used the results for two purposes: (1) to guide the search for effective MET distrubiton, which culminated in the "exponentially decaying MET" described in Section 2.3.2; and (2) to verify the correctness of our simulation results, since these DE equations precisely model the asymptotic behavior of the decoding process when the block size $n$ is large. We omit these DE equations here in the interest of space (as describing them needs considerable set-up and notations).

## 4 Evaluation

In this section, we evaluate our METTLE coding scheme (Section 2) under memoryless binary erasure channels (BECs) and Gilbert-Elliott (GE) channels in terms of coding efficiency and latency. Using different configurations, GE channels emulate real-world workloads such as VoIP (Voice over Internet Protocol), video streaming, and videoconferencing scenarios (Section 4.1.2 and Table 1). Here are the highlights of our evaluation results.

- Under BEC, METTLE achieves a 75% reduction in coding overhead compared to the baseline METTLE at 1% loss rate (Section 4.2.1).
- Under bursty channels (modeled by GE models), METTLE demonstrates stable coding efficiency and strong robustness against bursty or fast-fluctuating losses (Section 4.2.2).
- Compared with packet-level LDPC codes, METTLE attains slightly lower coding efficiency with significantly lower average decoding latency (Section 4.2.3).

- Under a BEC with a 1% loss rate, METTLE's average decoding latency is only 57 packets (Section 4.2.4).

## 4.1 Experimental Setup

We perform Monte Carlo simulations with 1000 independent runs per data point. Each run encodes $n = 10^5$ source packets into $(1+c)n$ codeword packets, where $c$ is the overhead ratio parameter. Codeword packets are then transmitted through a channel, where some packets will be lost. A run counts as successful only if all $n = 10^5$ source packets are successfully decoded by the receiver.

### 4.1.1 Performance Metrics.
We evaluate METTLE and its baseline schemes against the following metrics.

*Fraction of symbols undecoded (FSU).* FSU is the ratio between the number of unrecovered source symbols at the end of the decoding, and the total number of source symbols.

*Failure probability.* This is the probability that decoding fails to recover all source symbols from the received encoded packets. In practice, failure probabilities below around 1% are generally required to provide high quality of experience (QoE) [55].

*Overhead ratio (c).* The overhead ratio represents the relative excess number of codeword symbols required beyond the number of source symbols $n$, excluding the tail cost ratio, which can be reduced to $(1+c)\frac{w}{2n}$ (where $w$ denotes the coupling window size), as elaborated later.

*Latency.* This is the number of codeword symbols (packets) that must be received before a given source symbol can be recovered. To ensure fairness at different overhead levels, the latency is normalized by $(1+c)$, since the effective sending rate (and, equivalently, the bandwidth usage) scales linearly with this factor. Propagation delay is excluded since all schemes experience the same channel conditions.

### 4.1.2 Erasure Channels.
We evaluate the METTLE coding scheme under the following channels.

*Binary Erasure Channel (BEC).* The binary erasure channel is described by the parameter $\varepsilon$. It drops each packet independently with probability $\varepsilon$, as described in Section 3.1.

*Gilbert-Elliott (GE) channel.* A Gilbert-Elliott (GE) channel is described by parameters $(\epsilon, \delta, \alpha, \beta)$. GE channel [17], a.k.a. Markov-modulated BEC [58], is widely used for evaluating ECC performance under time-varying and/or bursty erasure conditions. The GE channel maintains a two-state (0-1) Markovian process. For each packet, if the current state is 0, it drops the packet with $\epsilon$ probability and transits to state 1 with $\alpha$ probability. Likewise, if the current state is 1, it drops the packet with $\delta$ probability and transits to state 0 with $\beta$ probability. As such, the overall packet loss probability is $(\alpha + \beta)^{-1}(\alpha\delta + \beta\epsilon)$. Conventionally, state 0 is called the "good" state, state 1 is called the "bad" state, and $\delta > \epsilon$. When $\delta = 1$, the channel experiences bursts of consecutive (packet) erasures. We evaluate multiple GE configurations to emulate a variety of network dynamics (e.g., VoIP, video streaming, and videoconferencing traces), as summarized in Table 1. Detailed trace characteristics and corresponding application scenarios are discussed later in Section 4.2.2.

### 4.1.3 Coding Schemes.
We evaluate four coding schemes, including our proposed METTLE and three other schemes. Their parameter settings are summarized below. The first three schemes (METTLE and its two baselines) are parameterized by the coding overhead ratio $c$, the time-coupling window size $w$, and the number of edges $l$ per ball (packet).

*METTLE.* This is the full-fledged METTLE with both MET and TLE. Each ball (at position $x$) has a TLE (from ball $x$ to bin $(1+c)x$). For each ball (at position) $x$, the landing positions ($\eta_1, \eta_2, \eta_3$ shown

Table 1. GE model configurations. The first row follows the setup in [5], the second uses the WiMAX model from [2], the third (light loss) and the fourth (heavy loss) correspond to the real video traces in [47], and the last row represents an extreme bursty case (average bad-state length = 100 packets).

| Configuration | $\alpha$ | $\beta$ | $\epsilon$ | $\delta$ | $p_{\text{avg}}$ |
|---|---|---|---|---|---|
| VoIP | $5 \times 10^{-4}$ | 0.2 | 0.01 | 1 | 1.25% |
| WiMAX video streaming | 0.04 | 0.05 | 0.01 | 0.02 | 1.44% |
| Videoconferencing-light | 0.05 | 0.75 | 0.01 | 0.1 | 1.56% |
| Videoconferencing-heavy | | | 0.05 | 0.5 | 7.81% |
| Long-burst | 0.001 | 0.01 | 0.01 | 0.1 | 1.82% |

in Fig. 1) of its $(l-1)$ non-TLE (MET) edges within its time-coupling window $[(1+c)x, (1+c)(x+w))$ are Binomial RVs with parameters $\zeta_2, \zeta_3, \cdots, \zeta_l$ respectively, as explained in Section 2.3.2. We start with subscript '2' in $\zeta$ because $x$'s TLE edge is the "first edge" by convention. As such, parameters $\zeta_2, \cdots, \zeta_l$ control the expected landing positions of the non-TLE edges, and the window size $w$ controls the trade-off between coding efficiency and latency. In our experiments, we use $l = 4$ with $(\zeta_2, \zeta_3, \zeta_4) = (1/2, 1/4, 1/8)$ (the "exponential decay" shown in Fig. 2). We choose $w = 600$, because our experiments on $w$ values within $[400, 1000]$ show that $w = 600$ strikes a good trade-off between coding efficiency and latency.

*Baseline METTLE (called baseline hereafter).* This is simply Walzer's scheme adapted to time-coupling and restricted to one-directional (left to right) peeling, but without either MET or TLE. For each ball (original packet) at position $x$, all its $l$ edges are i.i.d. uniformly distributed within $[(1+c)x, (1+c)(x+w))$, as described in Section 2.3.1. We set $l = 4$ and $w = 600$, which yield the best performance empirically.

*Baseline + TLE.* To study the benefit of TLE, we implement a variant of METTLE wherein each ball (at position $x$) has a TLE, and its $(l-1)$ non-TLE edges are i.i.d. uniformly distributed within $[(1+c)x, (1+c)(x+w))$. We set $l = 3$ (one TLE edge and two non-TLE edges) and $w = 600$ for best empirical performance. There is no MET in this variant.

*Packet-level LDPC.* For the reasons explained in Section 3.4 (first paragraph), we also compare METTLE with regular LDPC codes, as follows. The packet-level LDPC ensemble is constructed as a $(d_v, d_c)$-regular code, yielding an overhead ratio $c = d_v/(d_c - d_v)$. Each parity-check matrix $H \in \{0,1\}^{m \times n}$ has $n$ variable nodes (columns) and $m$ check nodes (rows), with exactly $d_v$ ones per column and $d_c$ ones per row. We generate $H$ by random socket matching between variable and check nodes while rejecting duplicate connections to keep $H_{ij} \in \{0, 1\}$. This randomized construction ensures uniform edge degree and serves as a baseline for evaluating METTLE's coding efficiency. Table 2 lists the $(d_v, d_c)$ choices used in our experiments, and the corresponding decoding thresholds are analyzed in Section 4.2.3.

## 4.2 Numerical Results

In this section, we present simulation results demonstrating the efficiency and latency performance of METTLE. Our goal is to highlight its potential and provide a foundation for future exploration, rather than exhaustively covering the entire design space. For clarity, we focus on representative settings instead of tuning all parameters for optimal performance. We view this study as an initial step toward a more comprehensive investigation of METTLE's variants, parameter choices, and performance boundaries.

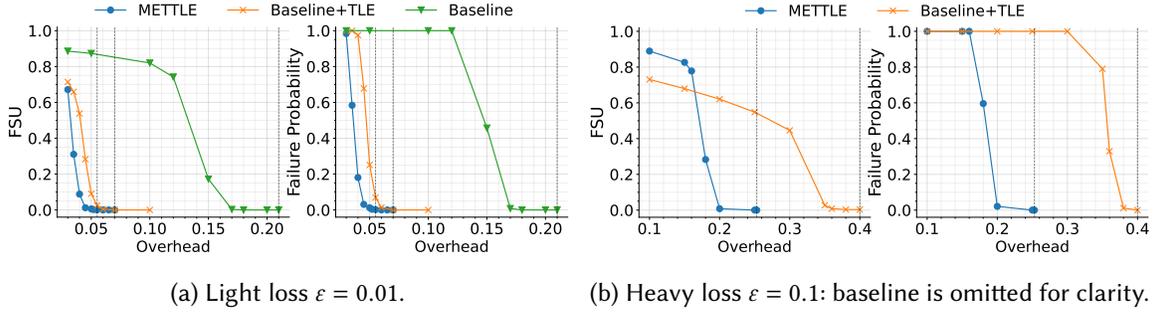

(a) Light loss $\varepsilon = 0.01$.

(b) Heavy loss $\varepsilon = 0.1$: baseline is omitted for clarity.

Fig. 3. Fraction of symbols undecoded (FSU) and failure probability vs. overhead under BEC.

*4.2.1 Coding Efficiency under BEC.* Fig. 3a shows the fraction of symbols undecoded (FSU) on the left (on y-axis) and failure probability on the right (on y-axis) versus the coding overhead (on x-axis), under a BEC with a random loss rate $\varepsilon = 0.01$. METTLE exhibits high coding efficiency, successfully recovering all symbols (failure $< 10^{-3}$) with an overhead below $c = 0.055$. In contrast, the baseline-only variant requires an overhead ratio of $c = 0.21$, and the baseline+TLE variant requires $c = 0.07$. This demonstrates that TLE alone yields a substantial coding efficiency gain compared to the baseline. The additional gain for METTLE arises from its MET component, which improves coding efficiency slightly.

As the erasure rate increases from $\varepsilon = 0.01$ (Fig. 3a) to $\varepsilon = 0.10$ (Fig. 3b), METTLE retains its efficiency at low overhead ratio, whereas the plots of the two baseline variants shift rightward, indicating a higher overhead requirement. We omit the baseline METTLE in the $\varepsilon = 0.1$ case, as it performs poorly under heavy loss and would clutter the figure without adding additional insight. This supports our claim that METTLE's design preserves high coding efficiency under harsher channel conditions.

Before evaluating METTLE's resilience under bursty losses, we address the *tail loss* issue that is illustrated in Fig. 1: the "east coast," or the "extra" coupling window at the end (whose length is $(1 + c)w$), is also a part of the coding overhead. This cost is similar to the rate loss caused by termination in SC-LDPC codes [28, 37, 48]. We did not include this tail loss in plotting Fig 3, because (1) it is small in our case (it is 0.6% when $n = 10^5$ and $w = 600$) and (2) all three schemes compared suffer from it equally and hence fairly. We will compress this tail using the technique to be shown next, and include the (compressed) tail cost in $c$ from this point on, for a fair comparison with LDPC.

The possibility of tail compression comes from the fact that, the "east east" is just as "thinly populated" as the "west coast," as explained in Section 2.2. Since METTLE cannot decode from right ("east") to left ("west") anyway, it is useless (for improving decoding success probability) to let the "east coast" be "thinly populated." Our idea is to shrink the size of the "east coast" by a factor just large enough to make the "east coast" as densely populated (in terms of the Poisson rate) as the "inland" region. To achieve this, we shrink the coupling range of the last $w$ balls by a factor that grows linearly from 1 (for the ball at position $n - w$) to 2 (for the ball at position $n - 1$). This adjustment compresses the tail by a factor of two while preserving overall coding efficiency as confirmed by our evaluation results. All subsequent evaluations use this tail compression technique.

*4.2.2 Coding Efficiency under Bursty Channels.* After evaluating performance under memoryless BEC channels, we next examine how METTLE performs in more realistic bursty-loss environments. Fig. 4 is "similarly structured" as Fig. 3. Fig. 4 shows performance results under GE models that emulate the bursty loss patterns observed in real-world traffic. The GE parameters, summarized in

Table 1, are derived from the traces statistics in [2, 5, 47]. The first row is the channel configuration following [5], which emulates VoIP-like bursty links with mean burst lengths of 5 packets. The second row uses the channel model from [2], simulating WiMAX wireless technology for video streaming. The third and fourth rows follow real-world video traces used in [47]. To illustrate an extreme case, the last configuration models a bursty channel where the bad state persists on average for 100 packets with an average of erasure rate 0.1, demonstrating performance limits under prolonged fades. Here, we compare METTLE's performance under bursty losses with its BEC results to demonstrate its robustness to temporal loss correlation.

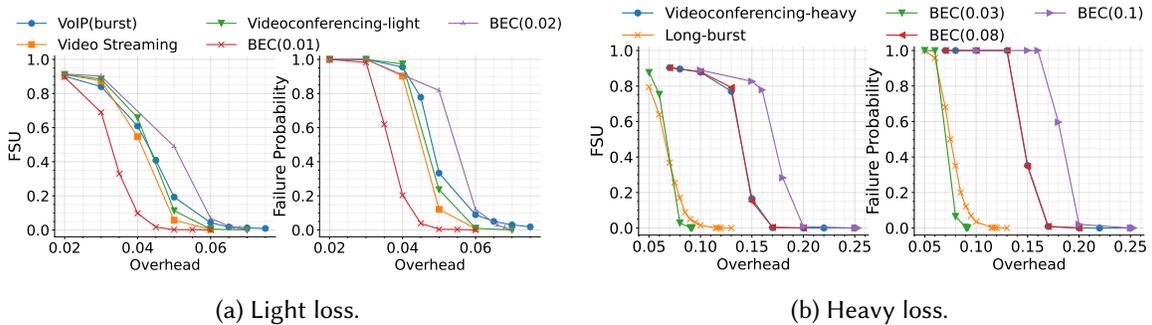

Fig. 4. Fraction of symbols undecoded (FSU) and failure probability vs. overhead under bursty channels.

Under bursty losses, the failure probability curves remain largely similar to those under i.i.d. losses (BEC). Specifically, as shown in Fig. 4a, for the three light-loss configurations (VoIP, WiMAX video streaming, and Videoconferencing-light, which have average loss rates 1.25%, 1.44%, and 1.56% respectively, as shown in Table 1), METTLE requires overheads slightly above the BEC(0.01) case but below the BEC(0.02) case (except in the case of VoIP that we will explain next). This result is consistent with our claim that METTLE only needs to "budget for (slightly above) the average." The VoIP configuration, which has an average burst length of five packets (longer than the typical two-packet case), exhibits a slightly longer failure-probability tail than BEC(0.02).

Under heavier losses (Videoconferencing-heavy and Long-burst in Fig. 4b), similar trends hold. The Long-burst configuration, where the bad state lasts longer (than the other configuration), requires only between 0.03 and 0.05 higher overhead ratio than that required for BEC with (Long-burst)'s average loss rate 1.82%, and performs far better than the $c = 0.253$ overhead typically needed to handle BEC with (Long-burst)'s worst-case loss rate of 10%. The plot of the Videoconferencing-heavy configuration (with average loss rate 7.81%), nearly overlaps with that of BEC(0.08) (i.i.d. loss at rate slightly larger than 7.81%). Overall, these results confirm that METTLE maintains strong resilience under bursty or fast-fluctuating channels, exhibiting high coding efficiency comparable to that under i.i.d. losses.

*4.2.3 Erasure Tolerance Comparison with LDPC.* In contrast to the previous coding efficiency analysis, which varies coding overhead $c$ at a fixed channel erasure probability $\varepsilon$, LDPC codes are constrained by their design: once the degree distribution or ensemble is chosen, the code rate is fixed. Consequently, their performance is typically reported in terms of the maximum erasure probability that can be sustained at a given rate. To align with this setting, we tune the rate of METTLE to match that of the LDPC ensembles and measure the maximum erasure probability for which reliable decoding (failure probability below $10^{-3}$) is achieved.

Table 2 compares the erasure tolerance of METTLE with that of regular LDPC (with block size $10^4$) ensembles of comparable rates.

Table 2. Erasure tolerance comparison between METTLE and LDPC ensembles. The empirical threshold for LDPC codes corresponds to a finite block length of $10^4$ symbols.

| Ensemble | Code Rate | Overhead Ratio($c$) | Empirical Threshold |
|---|---|---|---|
| METTLE | 0.95 | 5.263% | 0.01 |
|  | 0.9375 | 6.667% | 0.018 |
|  | 0.934 | 7.143% | 0.02 |
|  | 0.9 | 11.111% | 0.04 |
| (4,80) | 0.95 | 5.263% | 0.031 |
| (5,80) | 0.9375 | 6.667% | 0.036 |
| (6,90) | 0.934 | 7.143% | 0.036 |
| (4,40) | 0.9 | 11.111% | 0.068 |

As shown in Table 2, the empirical thresholds of METTLE is slightly lower than those of regular LDPC ensembles of comparable rate (after accounting for the tail loss $(w/2)/n = 0.3\%$ of METTLE, discussed in Section 4.2.1). For example, at a code rate of 0.9375 (overhead $\approx$ 6.7%), METTLE achieves a decoding threshold at $\varepsilon = 0.018$, while the (4, 80) LDPC ensemble shows a threshold of 0.036. Similarly, at a lower rate of 0.9 (overhead $\approx$ 11%), METTLE's empirical threshold ($\varepsilon = 0.04$) remains within a factor of two of the LDPC baseline ($\varepsilon = 0.068$). Overall, METTLE offers slightly lower erasure tolerance than regular LDPC codes, but achieves orders-of-magnitude lower decoding latency (LDPC's latency is $10^4$ packets in this case).

*4.2.4 Latency Performance.* We report detailed average latency results only for METTLE (under BEC), since other peeling-based methods (e.g., LT and its windowed variants) exhibit decoding delays on the order of encoding window size (which is at least tens of thousands of packets), far exceeding the low-latency target of our evaluation. As shown in Table 3, METTLE has low latency under both light and heavy losses.

Table 3. Latency performance under different erasure rates.

| Loss rate $\varepsilon$ | 0.01 | 0.02 | 0.03 | 0.1 |
|---|---|---|---|---|
| Latency (# of packets) | 57 | 84 | 117 | 199 |

*4.2.5 Girth Analysis.* As discussed in Section 2.3, METTLE can also be represented by a Tanner graph [54]. Therefore, girth analysis–a standard graph-theoretic tool used in LDPC code quality analysis–can also be applied to METTLE. Girth analysis (mentioned in Section 3.4) reports the graph's girth (its shortest cycle length) and the distribution of short-cycle lengths in the bipartite graph. Such statistics are an important indicator of an LDPC (or LDGM) code's quality: the existence of many short cycles is considered problematic. Because METTLE is a hashing-based code (Section 1.2.3), its bipartite graph is constructed randomly rather than through explicit graph optimization as in LDPC design. We can only control the edge-generation process (and distributions) across the different edge types. Hence, we analyze the distribution of different cycle lengths to make sure METTLE's hash-generated structure is not problematic.

For the girth analysis, we sample $10^4$ random edges from a randomly generated graph containing $10^5$ nodes, with an overhead ratio $c = 0.05$. For each sampled edge, we temporarily remove it

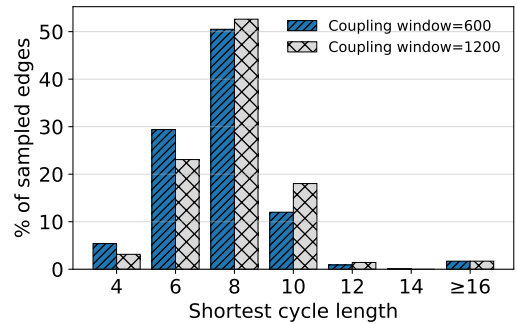

Fig. 5. Girth Distribution in METTLE.

and perform breadth-first search (BFS) from one endpoint to determine the shortest cycle length containing that edge.

Fig. 5 demonstrates cycle length distributions for two coupling window sizes: the default $w = 600$ and a larger $w = 1200$. As shown in Fig. 5, in the case of $w = 600$, most sampled edges participate in cycles of length 6-8, with the 25th percentile at length 6. Only less than 5% of the edges participate in cycles of length 4. This girth analysis result is considered reasonably good and shows that its deterministic TLE edges are not causing any "girth problem." The results in the case of $w = 1200$ is even better (e.g., with the percentage of length-4 cycles dropping to 3%).

## 5 Conclusion

In this work, we present METTLE, a new streaming erasure code that meet the requirements of three goals simultaneously in coding design: high efficiency, low complexity, and low latency. From spatial-coupling to time-coupling, and employing a multi-edge type (MET) with a deterministic touch-less leading edge (TLE) framework, METTLE achieves high coding efficiency while exhibiting the stream-decodable property. Our experimental results show that METTLE recovers all symbols with an overhead below 5.5% under 1% random loss, and maintains comparable efficiency to regular LDPC ensembles while offering much lower decoding latency. Under bursty or Markov-modulated channels, METTLE shows strong resilience, retaining performance close to that achieved under i.i.d. losses.

## References

[1] Bentaleb Abdelhak. 2024. Unifying real-time communications and content delivery with media-over-QUIC transport. https://events.vtools.ieee.org/m/406337.

[2] Laith Al-Jobouri, Martin Fleury, and Mohammed Ghanbari. 2013. Intra-Refresh Provision for WiMAX Data-Partitioned Video Streaming. *Consumer Electronics Times* 2, 3 (2013), 137–145.

[3] Congkai An, Huanhuan Zhang, Shibo Wang, Jingyang Kang, Anfu Zhou, Liang Liu, Huadong Ma, Zili Meng, Delei Ma, Yusheng Dong, and Xiaogang Lei. 2025. Tooth: toward optimal balance of video QoE and redundancy cost by fine-grained FEC in cloud gaming streaming. In *Proceedings of the 22nd USENIX Symposium on Networked Systems Design and Implementation* (Philadelphia, PA, USA) *(NSDI '25)*. USENIX Association, USA, Article 34, 17 pages.

[4] Ahmed Badr, Ashish Khisti, Wai-Tian Tan, and John Apostolopoulos. 2013. Streaming codes for channels with burst and isolated erasures. In *2013 Proceedings IEEE INFOCOM*. IEEE, Turin, Italy, 2850–2858. https://doi.org/10.1109/INFCOM.2013.6567095

[5] Ahmed Badr, Ashish Khisti, Wai-tian Tan, Xiaoqing Zhu, and John Apostolopoulos. 2017. FEC for VoIP using dual-delay streaming codes. In *IEEE INFOCOM 2017 - IEEE Conference on Computer Communications*. IEEE, Atlanta, GA, USA, 1–9. https://doi.org/10.1109/INFOCOM.2017.8057027

[6] Ahmed Badr, Pratik Patil, Ashish Khisti, Wai-Tian Tan, and John Apostolopoulos. 2016. Layered constructions for low-delay streaming codes. *IEEE Transactions on Information Theory* 63, 1 (2016), 111–141.

[7] Netflix Technology Blog. 2023. All of Netflix's HDR video streaming is now dynamically optimized. https://netflixtechblog.com/all-of-netflixs-hdr-video-streaming-is-now-dynamically-optimized-e9e0cb15f2ba

[8] Burton H. Bloom. 1970. Space/time trade-offs in hash coding with allowable errors. *Commun. ACM* 13, 7 (July 1970), 422–426. https://doi.org/10.1145/362686.362692

[9] Mattia C.O. Bogino, Pasquale Cataldi, Marco Grangetto, Enrico Magli, and Gabriella Olmo. 2007. Sliding-Window Digital Fountain Codes for Streaming of Multimedia Contents. In *2007 IEEE International Symposium on Circuits and Systems (ISCAS)*. IEEE, New Orleans, LA, USA, 3467–3470. https://doi.org/10.1109/ISCAS.2007.378373

[10] John W. Byers, Michael Luby, Michael Mitzenmacher, and Ashutosh Rege. 1998. A digital fountain approach to reliable distribution of bulk data. *ACM SIGCOMM Computer Communication Review* 28, 4 (1998), 56–67. Publisher: ACM New York, NY, USA.

[11] Pasquale Cataldi, Marco Grangetto, Tammam Tillo, Enrico Magli, and Gabriella Olmo. 2010. Sliding-Window Raptor Codes for Efficient Scalable Wireless Video Broadcasting With Unequal Loss Protection. *IEEE Transactions on Image Processing* 19, 6 (June 2010), 1491–1503.

[12] Jr. Daniel J. Costello, Min Zhu, David G. M. Mitchell, and Michael Lentmaier. 2025. High-Rate Spatially Coupled LDPC Codes Based on Massey's Convolutional Self-Orthogonal Codes. arXiv:2502.03774 [cs.IT] https://arxiv.org/abs/2502.03774

[13] Frederic Didier. 2009. Efficient erasure decoding of Reed-Solomon codes. arXiv:0901.1886 [cs.IT] https://arxiv.org/abs/0901.1886

[14] Martin Dietzfelbinger and Stefan Walzer. 2019. Dense Peelable Random Uniform Hypergraphs. arXiv:1907.04749 [cs.DS] https://arxiv.org/abs/1907.04749

[15] Yevgeniy Dodis, Leonid Reyzin, and Adam Smith. 2004. Fuzzy Extractors: How to Generate Strong Keys from Biometrics and Other Noisy Data. In *Advances in Cryptology - EUROCRYPT 2004*, Christian Cachin and Jan L. Camenisch (Eds.). Springer Berlin Heidelberg, Berlin, Heidelberg, 523–540.

[16] Elad Domanovitz, Silas L. Fong, and Ashish Khisti. 2022. An Explicit Rate-Optimal Streaming Code for Channels With Burst and Arbitrary Erasures. *IEEE Transactions on Information Theory* 68, 1 (2022), 47–65. https://doi.org/10.1109/TIT.2021.3121101

[17] Edwin O Elliott. 1963. Estimates of error rates for codes on burst-noise channels. *The Bell System Technical Journal* 42, 5 (1963), 1977–1997.

[18] David Eppstein, Michael T. Goodrich, Frank Uyeda, and George Varghese. 2011. What's the difference? efficient set reconciliation without prior context. *SIGCOMM Comput. Commun. Rev.* 41, 4 (Aug. 2011), 218–229. https://doi.org/10.1145/2043164.2018462

[19] A Jimenez Felstrom and Kamil Sh Zigangirov. 1999. Time-varying periodic convolutional codes with low-density parity-check matrix. *IEEE Transactions on Information Theory* 45, 6 (1999), 2181–2191.

[20] G Forney. 1971. Burst-correcting codes for the classic bursty channel. *IEEE Transactions on Communication Technology* 19, 5 (1971), 772–781.

[21] R. Gallager. 1962. Low-density parity-check codes. *IRE Transactions on Information Theory* 8, 1 (1962), 21–28. https://doi.org/10.1109/TIT.1962.1057683

[22] Michael T. Goodrich and Michael Mitzenmacher. 2011. Invertible bloom lookup tables. In *2011 49th Annual Allerton Conference on Communication, Control, and Computing (Allerton)*. IEEE, Monticello, IL, USA, 792–799. https://doi.org/10.1109/Allerton.2011.6120248

[23] Saqib Hakak, Thippa Reddy Gadekallu, Praveen Kumar Reddy Maddikunta, Swarna Priya Ramu, Parimala M, Chamitha De Alwis, and Madhusanka Liyanage. 2023. Autonomous vehicles in 5G and beyond: A survey. *Vehicular Communications* 39 (2023), 100551. https://doi.org/10.1016/j.vehcom.2022.100551

[24] Hemani Kaushal and Georges Kaddoum. 2016. Underwater Optical Wireless Communication. *IEEE Access* 4 (2016), 1518–1547. https://doi.org/10.1109/ACCESS.2016.2552538

[25] Tomer Keniagin, Eitan Yaakobi, and Ori Rottenstreich. 2025. CertainSync: Rateless Set Reconciliation with Certainty. *Proc. ACM Meas. Anal. Comput. Syst.* 9, 2, Article 18 (June 2025), 33 pages. https://doi.org/10.1145/3727110

[26] F. Krzakala, M. Mézard, F. Sausset, Y. F. Sun, and L. Zdeborová. 2012. Statistical-Physics-Based Reconstruction in Compressed Sensing. *Phys. Rev. X* 2 (May 2012), 021005. Issue 2. https://doi.org/10.1103/PhysRevX.2.021005

[27] Shrinivas Kudekar, Tom Richardson, and Rüdiger L. Urbanke. 2013. Spatially Coupled Ensembles Universally Achieve Capacity Under Belief Propagation. *IEEE Transactions on Information Theory* 59, 12 (2013), 7761–7813. https://doi.org/10.1109/TIT.2013.2280915

[28] Shrinivas Kudekar, Thomas J Richardson, and Rüdiger L Urbanke. 2011. Threshold saturation via spatial coupling: Why convolutional LDPC ensembles perform so well over the BEC. *IEEE transactions on information theory* 57, 2 (2011), 803–834.

[29] Shrinivas Kudekar, Thomas J. Richardson, and Rüdiger L. Urbanke. 2015. Wave-Like Solutions of General 1-D Spatially Coupled Systems. *IEEE Transactions on Information Theory* 61, 8 (2015), 4117–4157. https://doi.org/10.1109/TIT.2015.

2438870
[30] M. Luby. 2002. LT codes. In *The 43rd Annual IEEE Symposium on Foundations of Computer Science, 2002. Proceedings*. IEEE Comput. Soc, Vancouver, BC, Canada, 271–280. https://doi.org/10.1109/SFCS.2002.1181950
[31] Michael G. Luby, Michael Mitzenmacher, M. Amin Shokrollahi, Daniel A. Spielman, and Volker Stemann. 1997. Practical loss-resilient codes. In *Proceedings of the Twenty-Ninth Annual ACM Symposium on Theory of Computing* (El Paso, Texas, USA) *(STOC '97)*. Association for Computing Machinery, New York, NY, USA, 150–159. https://doi.org/10.1145/258533.258573
[32] Francisco Lázaro and Balázs Matuz. 2023. A Rate-Compatible Solution to the Set Reconciliation Problem. *IEEE Transactions on Communications* 71, 10 (2023), 5769–5782. https://doi.org/10.1109/TCOMM.2023.3296630
[33] D.J.C. MacKay. 1999. Good error-correcting codes based on very sparse matrices. *IEEE Transactions on Information Theory* 45, 2 (1999), 399–431. https://doi.org/10.1109/18.748992
[34] Emin Martinian and C-EW Sundberg. 2004. Burst erasure correction codes with low decoding delay. *IEEE Transactions on Information theory* 50, 10 (2004), 2494–2502.
[35] Emin Martinian and Mitchell Trott. 2007. Delay-Optimal Burst Erasure Code Construction. In *2007 IEEE International Symposium on Information Theory*. IEEE, Nice, France, 1006–1010. https://doi.org/10.1109/ISIT.2007.4557355
[36] François Michel, Quentin De Coninck, and Olivier Bonaventure. 2019. QUIC-FEC: Bringing the benefits of Forward Erasure Correction to QUIC. In *2019 IFIP Networking Conference (IFIP Networking)*. IEEE, Warsaw, Poland, 1–9. https://doi.org/10.23919/IFIPNetworking.2019.8816838
[37] David GM Mitchell, Michael Lentmaier, and Daniel J Costello. 2015. Spatially coupled LDPC codes constructed from protographs. *IEEE Transactions on Information Theory* 61, 9 (2015), 4866–4889.
[38] David G. M. Mitchell, Ali E. Pusane, Kamil Sh. Zigangirov, and Daniel J. Costello. 2008. Asymptotically good LDPC convolutional codes based on protographs. In *2008 IEEE International Symposium on Information Theory*. IEEE, Toronto, ON, Canada, 1030–1034. https://doi.org/10.1109/ISIT.2008.4595143
[39] Thiago Camelo Mourão, Shady Saikali, Evan Patel, Mischa Dohler, Vipul Patel, and Márcio Covas Moschovas. 2025. Chapter 15 - Telesurgery applications, current status, and future perspectives in technologies and ethics. In *Handbook of Robotic Surgery*, Stênio de Cássio Zequi and Hongliang Ren (Eds.). Academic Press, 161–168. https://doi.org/10.1016/B978-0-443-13271-1.00027-3
[40] Naruomi Obata, Yung-Yih Jian, Kenta Kasai, and Henry D. Pfister. 2013. Spatially-coupled multi-edge type LDPC codes with bounded degrees that achieve capacity on the BEC under BP decoding. In *2013 IEEE International Symposium on Information Theory*. IEEE, Istanbul, Turkey, 2433–2437. https://doi.org/10.1109/ISIT.2013.6620663
[41] Leonardo Peroni and Sergey Gorinsky. 2025. An End-to-End Pipeline Perspective on Video Streaming in Best-Effort Networks: A Survey and Tutorial. *ACM Comput. Surv.* 57, 12, Article 322 (July 2025), 47 pages. https://doi.org/10.1145/3742472
[42] I. S. Reed and G. Solomon. 1960. Polynomial Codes Over Certain Finite Fields. *J. Soc. Indust. Appl. Math.* 8, 2 (1960), 300–304. https://doi.org/10.1137/0108018
[43] Tom Richardson and Ruediger Urbanke. 2008. *Modern Coding Theory*. Cambridge University Press, Cambridge. https://doi.org/10.1017/CBO9780511791338
[44] Thomas J Richardson and Rüdiger L Urbanke. 2002. The capacity of low-density parity-check codes under message-passing decoding. *IEEE Transactions on information theory* 47, 2 (2002), 599–618.
[45] Michael Rink. 2013. Mixed Hypergraphs for Linear-Time Construction of Denser Hashing-Based Data Structures. In *SOFSEM 2013: Theory and Practice of Computer Science*, Peter van Emde Boas, Frans C. A. Groen, Giuseppe F. Italiano, Jerzy Nawrocki, and Harald Sack (Eds.). Springer Berlin Heidelberg, Berlin, Heidelberg, 356–368.
[46] Michael Rudow and KV Rashmi. 2022. Streaming codes for variable-size messages. *IEEE Transactions on Information Theory* 68, 9 (2022), 5823–5849.
[47] Michael Rudow, Francis Y. Yan, Abhishek Kumar, Ganesh Ananthanarayanan, Martin Ellis, and K.V. Rashmi. 2023. Tambur: Efficient loss recovery for videoconferencing via streaming codes. In *20th USENIX Symposium on Networked Systems Design and Implementation (NSDI 23)*. USENIX Association, Boston, MA, 953–971. https://www.usenix.org/conference/nsdi23/presentation/rudow
[48] Laurent Schmalen. 2015. Spatially Coupled LDPC Codes – Theory and Applications. Tutorial slides, Karlsruhe Institute of Technology (KIT). https://www.cel.kit.edu/english/370.php.
[49] Amin Shokrollahi. 2006. Raptor codes. *IEEE Transactions on Information Theory* 52, 6 (2006), 2551–2567. https://doi.org/10.1109/TIT.2006.874390
[50] Amin Shokrollahi and Michael Luby. 2011. *Raptor codes*. Vol. 6. Now Publishers, Inc. 213–322 pages. https://doi.org/10.1561/0100000060
[51] A. Shokrollahi, M. Luby, M. Watson, T. Stockhammer, and L. Minder. 2011. RaptorQ Forward Error Correction Scheme for Object Delivery. RFC 6330. Available at https://datatracker.ietf.org/doc/html/rfc6330.

[52] R. Singleton. 1964. Maximum distanceq-nary codes. *IEEE Transactions on Information Theory* 10, 2 (1964), 116–118. https://doi.org/10.1109/TIT.1964.1053661

[53] B. Sklar. 1997. Rayleigh fading channels in mobile digital communication systems. I. Characterization. *IEEE Communications Magazine* 35, 9 (1997), 136–146. https://doi.org/10.1109/35.620535

[54] R. Tanner. 1981. A recursive approach to low complexity codes. *IEEE Transactions on Information Theory* 27, 5 (1981), 533–547. https://doi.org/10.1109/TIT.1981.1056404

[55] International Telecommunication Union. 2001. *ITU-T Recommendation G.1010: End-user multimedia QoS categories*. Technical Report G.1010 (11/2001). ITU-T. https://www.itu.int/rec/T-REC-G.1010-200111-I

[56] Stefan Walzer. 2025. Peeling Close to the Orientability Threshold Spatial Coupling in Hashing-Based Data Structures. *ACM Trans. Algorithms* 21, 3, Article 33 (July 2025), 23 pages. https://doi.org/10.1145/3711822

[57] Lei Yang, Yossi Gilad, and Mohammad Alizadeh. 2024. Practical Rateless Set Reconciliation. In *Proceedings of the ACM SIGCOMM 2024 Conference* (Sydney, NSW, Australia) *(ACM SIGCOMM '24)*. Association for Computing Machinery, New York, NY, USA, 595–612. https://doi.org/10.1145/3651890.3672219

[58] Yang Yang, Jian Tan, Ness B. Shroff, and Hesham El Gamal. 2014. Delay Asymptotics With Retransmissions and Incremental Redundancy Codes Over Erasure Channels. *IEEE Transactions on Information Theory* 60, 3 (2014), 1932–1944. https://doi.org/10.1109/TIT.2014.2300485

[59] Arvind Yedla, Yung-Yih Jian, Phong S. Nguyen, and Henry D. Pfister. 2012. A simple proof of threshold saturation for coupled scalar recursions. In *2012 7th International Symposium on Turbo Codes and Iterative Information Processing (ISTC)*. IEEE, Gothenburg, Sweden, 51–55. https://doi.org/10.1109/ISTC.2012.6325197

[60] Arvind Yedla, Yung-Yih Jian, Phong S. Nguyen, and Henry D. Pfister. 2012. A simple proof of threshold saturation for coupled vector recursions. In *2012 IEEE Information Theory Workshop*. IEEE, Lausanne, Switzerland, 25–29. https://doi.org/10.1109/ITW.2012.6404671


\end{document}